\documentclass[%
 reprint,
amsmath,amssymb,
aps,
]{revtex4-2}

\usepackage{graphicx}% Include figure files
\usepackage{dcolumn}% Align table columns on decimal point
\usepackage{bm}% bold math
\usepackage[usenames]{color}

\makeatletter
\newenvironment{figurehere}
{\def\@captype{figure}}
{}
\makeatother

\begin{document}

%\preprint{APS/123-QED}

\title{The infuence of compressive lattice deformations on the zone-center energy band properties of zincblende GaN and InN. Hybrid density functional results\\}% Force line breaks with \\

\author{Juli\'an David Correa}
\affiliation{%
 Facultad de Ciencias B\'asicas. Universidad de Medell\'{\i}n. Medell\'{\i}n, Colombia.
}%


\author{Miguel Eduardo Mora-Ramos}
\email{memora@uaem.mx}
\affiliation{
 Centro de Investigaci\'on en Ciencias-IICBA. Universidad Aut\'onoma del Estado de Morelos. Av. Universidad 1001, C.P. 62209, Cuernavaca, Morelos, Mexico.\\
}%
%\affiliation{
%Third institution, the second for Charlie Author
%\affiliation{%


\date{\today}% It is always \today, today,

\begin{abstract}
We have investigated the effects of hydrostatic pressure and compressive biaxial strain on the $\Gamma$-point energy states of GaN and InN with zincblende crystal structure via first-principles DFT+HSE06 computation. To correctly reproduce accepted experimental values of the energy band gap of both compounds, the procedure includes modified exchange-correlation fractions in the HSE hybrid functional, changing from the standard value $\alpha=0.25$ to $\alpha=0.43$ (GaN) and $\alpha=0.40$ (InN). Within this environment, the work reports on the variation of conduction and valence band edges as functions of the lattice deformation. In addition, we present fitted expressions describing the change in the band gap and the spin-orbit splitting energy due to unit cell size modification. All these parameters are main input quantities in the description of electronic and optical properties of InN/GaN-based heterostructures.
%\begin{description}
%\end{description}
\end{abstract}

\maketitle

%\tableofcontents

\section{\label{intro}Introduction
%First-level heading:\protect\\ The line break was forced \lowercase{via} \textbackslash\textbackslash
}

III-V nitride semiconductors have acquired relevant technological significance after their succesful application -mainly in their nanostructured forms- to the design and development of optoelectronic devices in the visible-to-ultraviolet range of the electromagnetic spectrum as well as due to their usage in high-carrier-mobility electronic devices \cite{Chen2020, Asbeck2019,As2018}. The natural appearance of these compounds corresponds to a crystalline structure of the wurtzite (WZ, hexagonal) type, which reveals as their lowest-enthalpy allotropic form. However, the zincblende (ZB, cubic) phase is only slightly higher in energy, and it has been possible to grow it in the form of thin films with suitable thermodynamical stability \cite{Deppe2020,Baron2020, Baron2019,Lee2018,As2013, Novikov2011,Novikov2010, As2006,Wang2000,Sitar1989}. Contrary to the wurtzite phase, the zincblende -as a centrosymmetric material- lacks internal electric polarizations in the unit cell; which has been named as the source of emission efficiency loss when using WZ III-V nitrides as core elements of light-emitting diodes and semiconductor lasers (see \cite{Lee2017,laserfocus2016} and references therein). So, in the particular case of those application lines, it has become desirable to gradually switch towards the use of cubic nitride materials. This is also a tendency for the development of other kinds of low-dimensional semiconductor structures based on III-V nitrides \cite{As2014,Mietze2013,As2009,Schormann2006,Li2005A,Li2005B,Leite2002}. 

Due precisely to the interest in cubic GaN-InN heterostructures, the issue of the influence of biaxial crystal strain on the optical and electronic properties is of paramount importance. Since the values of the lattice constant of both materials differ in approximately 7\%, the InN-based quantum well regions become compressively deformed. Certainly, this phenomenon mainly manifest when layers are thin enough due to the appearance of lattice threading defects that induce strain relaxation in thicker layers, thus giving rise to a critical thickness value which, normally, implies very few monolayers. However, the use of metamorphic growing would allow to produce strained InGaN/GaN with higher values of In concentration and larger widths of strained well regions. At this point, it is worth mentioning a recent study by Ohtake \textit{et al.} on ZB InAs grown onto GaAs substrate by heteroepitaxy \cite{Ohtake2020}. These compounds have a difference in lattice constants that amounts 7.2\%, with InAs as the one that undergoes a biaxial compression. In their experimental work, the authors show that although a fully InAs strained layer would be only 2 monolayers thick, the InAs region will show only a partial relaxation even for a width of 100 nm. This result makes relevant the need to take strain effects into account when describing the energy structure of mismatched ZB III-V heterostructures and, could be possibly extended to deal with cubic-nitride-based ones. On the other hand, it is well known that uniaxial strain plays a significant role in the case of WZ nitride nanosystems. This has been recently verified through experimental investigation of InGaN/GaN quantum wells in which uniaxial and hydrostatic deformations wer induced \cite{Bercha2019, Bercha2018}. Throughout the years, theoretical research on strain effects in both ZB and WZ nitrides have been carried out with the use of microscopic approaches such as density-functional theory (DFT) and empirical pseudopotentials \cite{Tang2020,Karakostas2017,Oliva2012,Merad2004, Zunger2002, Tadjer1999}. 

With regard to the DFT calculations, different works reporting on the electronic and structural properties of III-nitrides have appeared through the years both investigating zincblende \cite{Oukli2021,Moussa2018,Araujo2013} and wurtzite \cite{Suski2009, Deligoz2007}. Hybrid exchange-correlation functionals of the type proposed by Heid, Scuseria and Ernzenhoff (HSE) \cite{HSE} have been used as well, with studies on the fundamental properties of zincblende \cite{Bayram2020,Landman2013,Mietze2011} and wurtzite III-V \cite{Bayram2020,Karakostas2016,Rinke2009} nitride semiconductors. Particularly, the most recent report by Tsai and Bayram explicitly report on the use of modified hybrid HSE functionals to suitably evaluate the band alignments of ternary nitride compounds. Another treatment which includes modification of HSE parameters to deal with cubic nitride structures was given in \cite{Landman2013}. This approach can be traced back to the work by Wadhera and coworkers \cite{Wadhera2010}, in which the modification of the exact Fock exchange percentage in the hybrid functional allowed to exactly fit the experimental energy band gap of zincblende GaAs. Among these works, the effect of lattice strain on the electronic of nitrides has been considered in \cite{Rinke2009}. There, the authors present results for deformation potentials for AlN, GaN, and InN. However, when speaking about hydrostatic deformation, the reports of pressure effects on the properties of these materials are  more widely present in the literature. 

Hydrostatic pressure (HP) has traditionally revealed as a tool to expose the features of electron and phonon properties in solids, as well as to point at the formation of new crystalline phases of a given material. Application of pressure to certain compounds became a rather spectacular event when room-temperature superconductivity was achieved \cite{Flores2020,Snider2020}. In the case of the III-N semiconductors of our interest, first-principles DFT investigations have appeared considering zincblende materials \cite{Hattabi2019,Usman2013,Riane2010,Li2005,Merad2003}  as well as wurtzite ones \cite{Li2005,Duan2015,Said2012,Peng2009}. In the latter case, the work by Duan \textit{et al.} have included HSE+GW calculations in order to determine the energy band gap of InN with greater accuracy. These authors also conclude that the wurtzite phase in this material is stable up to rather high values of HP \cite{Duan2015}.

The phenomenon of a pressure-induced structural phase transition in III-V nitrides has been one of the most extensively treated in what has to do with the effect of HP. Experimental research on the subject appears commented in the works \cite{Oliva2014, Oliva2013,Ibanez2012}, whereas a much extensive literature has been devoted to its theoretical investigation (mostly through the use of DFT-based approaches) \cite{Daoud2018,Kunc2015,Schwarz2014,Saoud2011,Verma2010,Duan2010,Cheng2007,Silva2005, Serrano2000,Pandey1993,Gorczyca1993}. In general, it is accepted that nitrides evolve from wurtzite or zincblende phases to rocksalt one, under high enough pressure values.

Cubic III-V nitrides GaN and InN are suitable constituents of low-dimensional heterostructures which can be the core of prospective electronic and optoelectronic devices. For that reason, the theoretical exploration of their electronic properties under lattice deformation would provide essential data for determining the charge carrier states in such nanosystems. This is due to the fact that  given the significant difference in their lattice constants crystal strain appears at the time of growing heterolayers based on both materials or their alloys. On the other hand, HP can be used as an external tool to influence on the carrier states in the system, with the consequent changes in the related properties. 

The most widely used tool to calculate energies and wavefunctions of electrons and holes in quantum wells (QWs) and other heterostructures is the $\vec{k}\cdot\vec{p}$ theory (see, for instance \cite{Rejeb2011}). This method requires the use of several input quantities (energy gaps, band off-sets, effective masses, band nonparabolicity parameters) associated to each building material in the structure. When strain (or HP) is taken into account, the corresponding variations of such data with lattice deformation are required in order to provide better quantitative estimations. Up to our knowledge, this information is not currently available for the above mentioned materials. For that reason, here we carry out a calculation of zone-center band structure in zincblende GaN and InN, focusing on the energy states directly above and below the band gap. The aim is to derive expressions for the band edge positions, energy band gap and conduction effective mass, as functions of two types of compressive lattice deformations: biaxial strain and hydrostatic pressure. For that purpose, we use a first-principles DFT+HSE06 approach. The article is organized as follows. Section 2 contains some details of computational work. In section 3 we present and discuss the obtained results and, in section 4 the conclusions of the work are given.

\section{\label{metod}Details of computational work}

It is known that first-principles calculations to determine the electronic structure of the semiconductors, using DFT+LDA (Local Density Approximation) or DFT+GGA (Generalized Gradient Approximation), have been characterized by a notorious underestimation of the value of the energy band gap, $E_g$. It turns out that this quantity is of paramount importance when dealing with the electronic and optical properties of these materials and, as well as when they are used to built low-dimensional heterostructures. Therefore, it becomes necessary resorting to a more reliable determination. Since our purpose is to describe as accurately as possible the effect of biaxial stress on the energy spectrum of c-GaN and c-InN, the chosen approach will be DFT+HSE, which uses functional exchange and correlation hybrids.

In 2004-2006, Heyd, Scuseria and Ernzerhof \cite{HSE} proposed a particular way of writing the Hartree-Fock (HF) functional, which contains the exchange, $E_x$, and correlation, $E_c$, interactions. This made it possible to significantly improve the results for $E_g$ and to bring them closer to the known values of the experiments. In this modification, the nucleus of the Coulombic interaction, $1/r$, in the contribution $E_x$ is separated into the sum of a short-range part (SR) and a long-range part (LR) with: \vspace{-0.5cm}

\begin{eqnarray}
	(SR,hyb): &\;\; &\frac{erfc(\mu r)}{r};\nonumber\\
	(LR,hyb): &\;\;  & \frac{erf(\mu r)}{r}.
\end{eqnarray}

In these expressions,  $er\!f(x)$ and $er\!fc(x)$ represent the ``error" and ``error-complement" functions, respectively. So, the exchange and correlation hybrid functional (HSE06) that is used as input in the calculation consists of a weighted distribution of the SR-contribution, with respect to the one usually used through the PBE (Purdue-Burnke-Ernzerhoff) functional. Meanwhile, the LR and correlation parts are kept in the form provided by the latter functional \cite{Landman2013}. Accordingly, it is possible to write: 
\vspace{-0.3cm}

\begin{eqnarray}
	E_{xc}^{hyb}(\mu) &= &\alpha E_x^{(SR,hyb)}(\mu)+(1-\alpha)E_x^{(SR,PBE)}(\mu)+\nonumber \\
	&  &+E_x^{(LR,PBE)}(\mu)+E_c^{PBE}.
\end{eqnarray}

The $\alpha$ parameter represents the exact Fock exchange percentage. In the original definitions, this parameter is assigned the value $\alpha=0.25$, while the value $\mu=0.2\ $\AA$^{-1}$ \cite{HSE} is set. However, it is often the case, particularly, when dealing with wide gap materials, that optimized values of these parameters are sought. For example, in \cite{Mietze2011} it remains $\alpha=0.25$ but $\mu$ varies between $0.2\,$\AA$^{-1}$ and $0.3\,$\AA$^{-1}$, to calculate the band offset of a c-GaN/c-AlN heterojunction. But in the case of III-V arsenides with zincblende structure, the correct representation of the GaAs and AlAs gaps is obtained using $\alpha=0.30$, while for InAs the value that adjusts the experimental result is $\alpha\sim0.28$ \cite{Wadhera2010}.

In general, the use of HSE-type hybrid functionals has shown significant improvements in the description of electronic gaps, and the relative alignments of bands (band offsets) in semiconductor heterostructures, as well as of the levels of defects, together with the exact prediction of transitions from direct to indirect gap in different materials. However, the gain in accuracy brings with it a large increase in computational cost of two orders of magnitude compared to that of conventional DFT treatment \cite{Landman2013}.

In our case, the accepted experimental gap of c-GaN could be reproduced quite accurately using a higher value of the parameter of exchange percentage, $\alpha=0.43$. This might happen due to the fact that the gap of this semiconductor compound is noticeably greater than in III-arsenides. However, in the case of c-InN the accepted energy gap value is reproduced using $\alpha=0.40$. So, the reason could also be related with the kind o anion atom we are dealing with.  Then considering these results, we proceeded to calculate the electronic structure around the center of the Brillouin zone. The reason for restricting to this region is that, as we said,  information about quantities such as the (direct) gap and the effective masses is essential for a description of the spectrum of III-V semiconductor-based heterostructures using the $\vec{k}\cdot\vec{p}$ approximation at $\Gamma$ point. In fact, determining these quantities, under the effect of strain, is an objective of this study. The \textit{ab-initio} computational tool employed in this work is the \textit{Quantum Espresso} code. We name the procedure as DFT+HSE06(mod). It starts with a DFT+GGA calculation to determine the equilibrium lattice constant $a$ for each compound investigated. The wave functions were expanded in a basis set of plane waves employing an energy cutoff of  80 Ry. For the convergence of total energy, a  k-grid of  $12\times 12\times 12$ points was employed. The geometric structure was relaxed until each atom's atomic force was less than $10^-4\,$Ry/Bohr. The lattice constant was reduced by  compression biaxial stress and fixed its value in the $xy$-plane, and the $a_z$ component was free in the relaxed processes. In the case of hydrostatic compression, $a$ reduces in a homogeneous way along all spatial directions, thus reducing the cell volume. In all cases, for relaxed structure, the corrections due to spin-orbit interaction are included. For the HSE06  calculation, an energy cutoff in the Fock operator of $130\,$Ry was used, together with a grid of $6\times 6 \times 6$ points.  The HSE06(mod) band structure was then obtained with the Wannier90 package \cite{Wannier2014}.

%Another important aspect that has limited us to work near the center of the Brillouin zone is that, apparently, there is still no version of \textit {Quantum Espresso + Wannier 90} that allows to correctly generate the energy bands  throughout the entire zone, using the DFT+HSE results, when the strain values ??are increasing. However, we have verified that it does without problems for HSE+GGA. We also check that the GGA-related calculation of the effective masses in $\Gamma$ does correctly reproduce experimental results. This has allowed us to use this simpler approach for that evaluation, without losing accuracy, when introducing the stress effects.

\section{Results}

Here we present the outcome of the DFT+HSE06(mod) first-principles calculation for electron states of cubic GaN and InN, at the Brillouin zone center, affected by two kinds of compressive lattice distortions: biaxial strain perpendicular to $[001]$ crystal direction, and hydrostatic deformation. The former usually manifests in heterostructures with InGaN layers grown onto GaN substrates. For these materials, GGA structural optimization gives $a_{GaN}=4.55\,$\AA, and $a_{InN}=5.03\,$\AA for the undeformed lattice constants. Comparing such results with those commonly accepted   ($a_{GaN}=4.50\,$\AA,  $a_{InN}=4.98\,$\AA \cite{Rinke2008}), it is possible to identify a fairly good accordance, bearing in mind the tendency of GGA to overestimate the value of distances. Since the analysis of deformation is made in relative terms, the difference in question should not have a noticeable impact on the discussion below . 

Firstly, we shall present the results concerning the characteristic energies at $k=0$ (lowest conduction band and upper valence band states) both for the biaxial strain and hydrostatic compressive deformations. In the strained case, energy values appear as functions of $\sigma=\delta a/a$, where $\delta a=a_{def}-a$ with $a_{def}$ corresponding to the compressed $xy$ lattice constant value. Furthermore, it is worth mentioning that overall zeroth reference for energies is taken at the unstrained upper valence band top. 

When dealing with the -compressive- hydrostatic deformation, the calculated results or the band parameters will appear as functions of the absolute value of the homogeneous relatve variation of the lattice constant, $\varepsilon=\vert \delta a/a\vert$, occurring along the three spatial directions in the crystal. That is, such a compression implies a relative variation of the unit cell volume in $\delta V/V=(a-\delta a)^3/a^3$.

\vspace{-1.0cm}

\subsection{The case of strained ZB GaN}

Figures 1-3 contain, respectively, the variation of the energy position of the zone-center edges of conduction ($c$), heavy-hole ($v1$), and light-hole ($v2$) bands. Actually, the quantities plotted are the corrections due to the strain effect, measured with respect to the corresponding values without deformation.

\vspace{0.3cm}
\begin{figurehere}
	\centering
	\includegraphics[width=7.5cm]{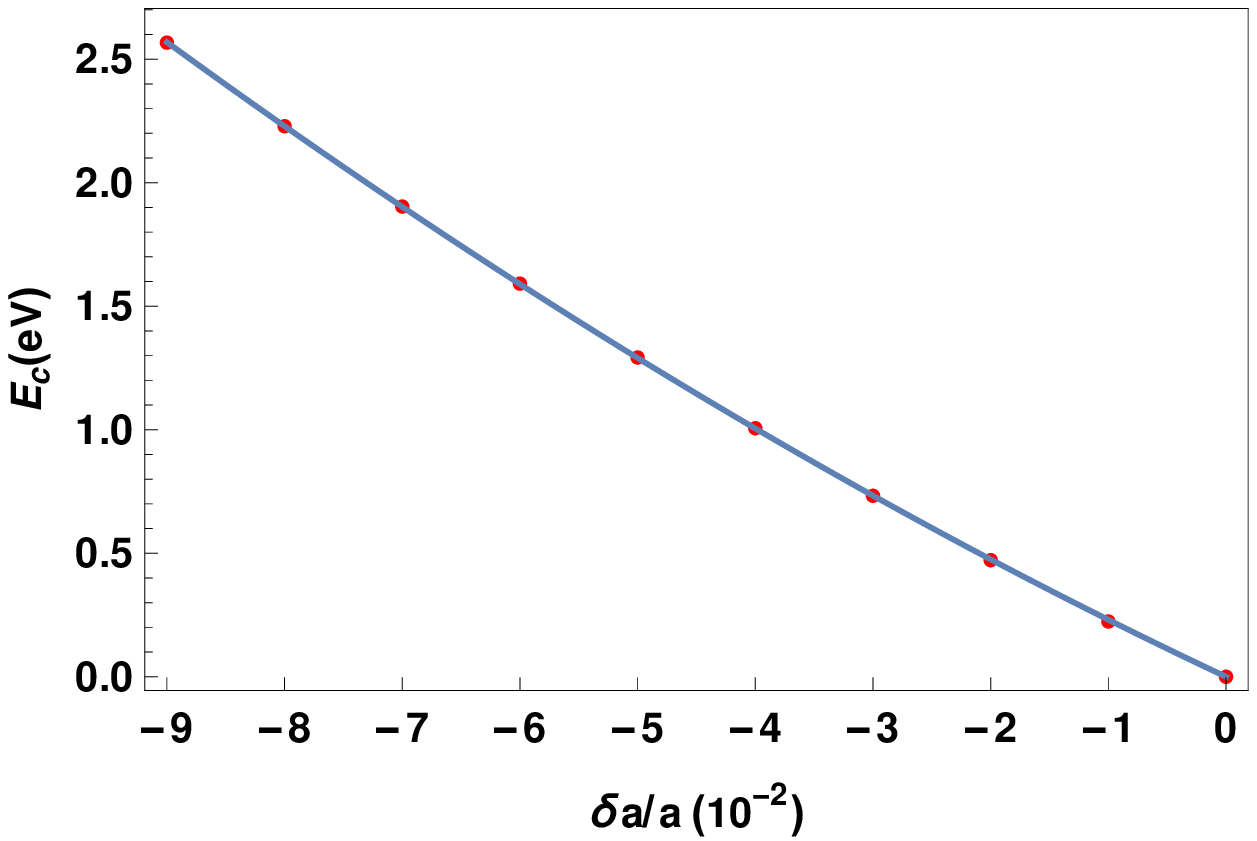}
	\caption{Calculated correction to the zone center energy of the conduction band bottom edge in zincblende GaN as a function of biaxial strain, measured from the unstrained value.}
	\label{FIG1}
\end{figurehere}

\vspace{0.3cm}
\begin{figurehere}
	\centering
	\includegraphics[width=7.5cm]{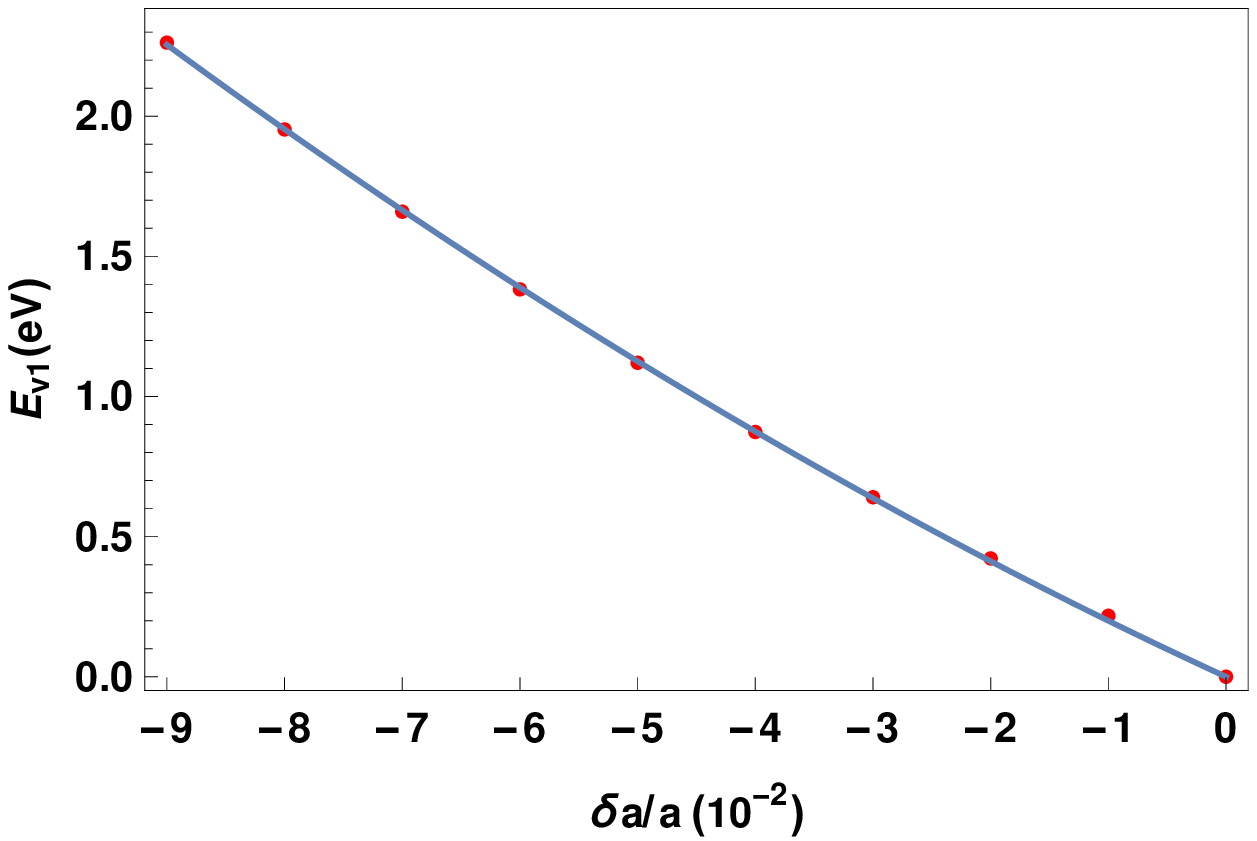}
	\caption{Calculated correction to the zone center energy of the heavy hole valence band ($v1$) top edge in zincblende GaN as a function of biaxial strain, measured from the unstrained value, which is taken as the zero energy reference in this work.}
	\label{FIG2}
\end{figurehere}

Our calculation has allowed to extend the range of biaxial compression until reaching a value of the in-plane lattice constant equal to $91\%$ of the unstrained one. It is possible to observe that the influence of strain makes the band edges to increase in their energy positions, and this effect is more pronounced in the case of conduction band bottom. In all cases, the solid lines correspond to best polynomials fittings for the  obtained results. Such a growing trend of the band edge energy positions agrees with that reported by us for zincblende GaAs, calculated using the tight-binding approach \cite{Mozo2021}.

\vspace{0.3cm}
\begin{figurehere}
	\centering
	\includegraphics[width=7.5cm]{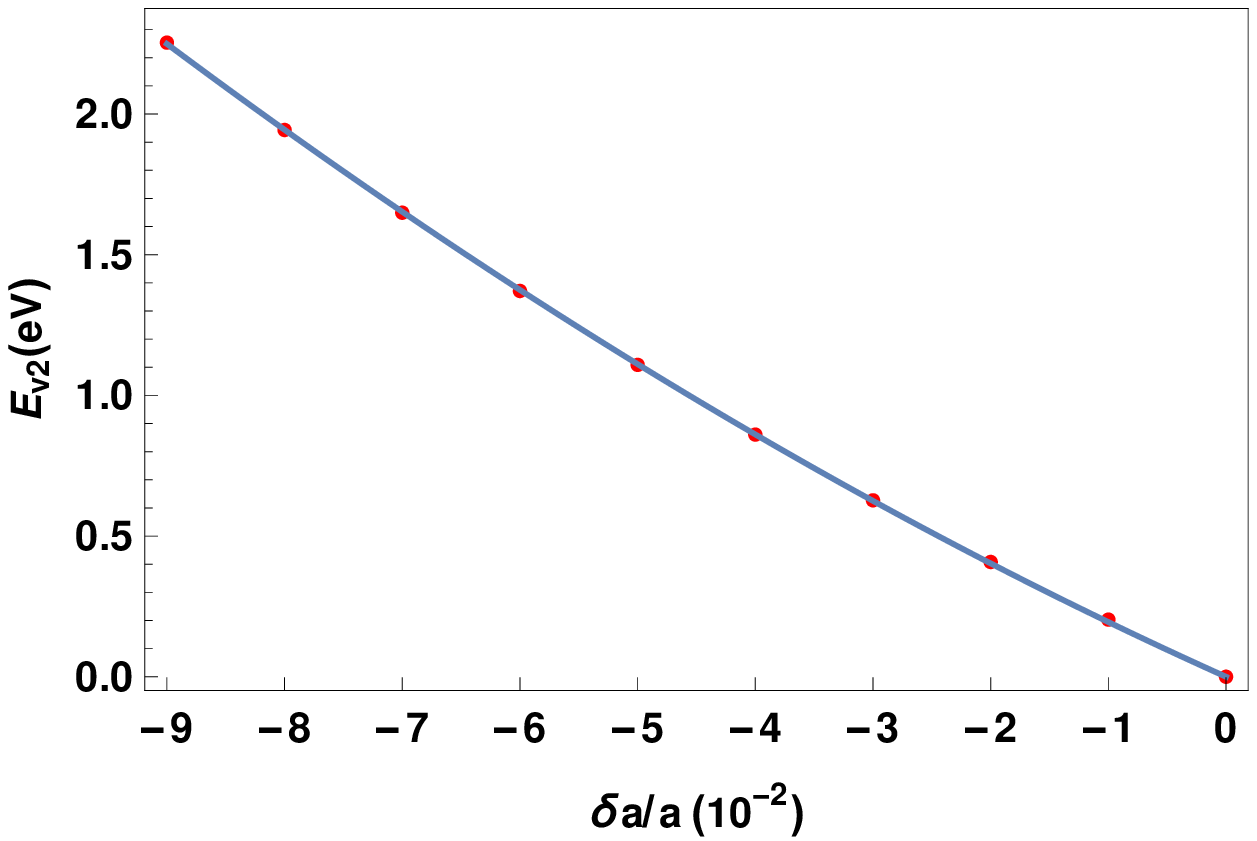}
	\caption{Calculated correction to the zone center energy of the light hole valence band ($v2$) top edge in zincblende GaN as a function of biaxial strain, measured from the unstrained value.}
	\label{FIG3}
\end{figurehere}

Since the variation for these energy positions is not homogeneous, the direct band gap  ($E_g$) at $k=0$ also augments as a consequence of the compressive biaxial strain, as observed from Fig. 4. The DFT+HSE06(mod) result, with $\alpha=0.43$, for the unstrained gap value is $E_g(0)=3.3326\,$eV. This result is in very good agreement with accepted one reported in Ref. \cite{Rinke2008}. A strain value of $9\%$ causes the increment of about $0.3\,$eV in the band gap. This is a rather moderate growth which allows us to predict that biaxially Compressed ZB GaN would remain as a direct gap material within the range of deformation considered. This statement is done taking into account 

\vspace{0.3cm}
\begin{figurehere}
	\centering
	\includegraphics[width=7.5cm]{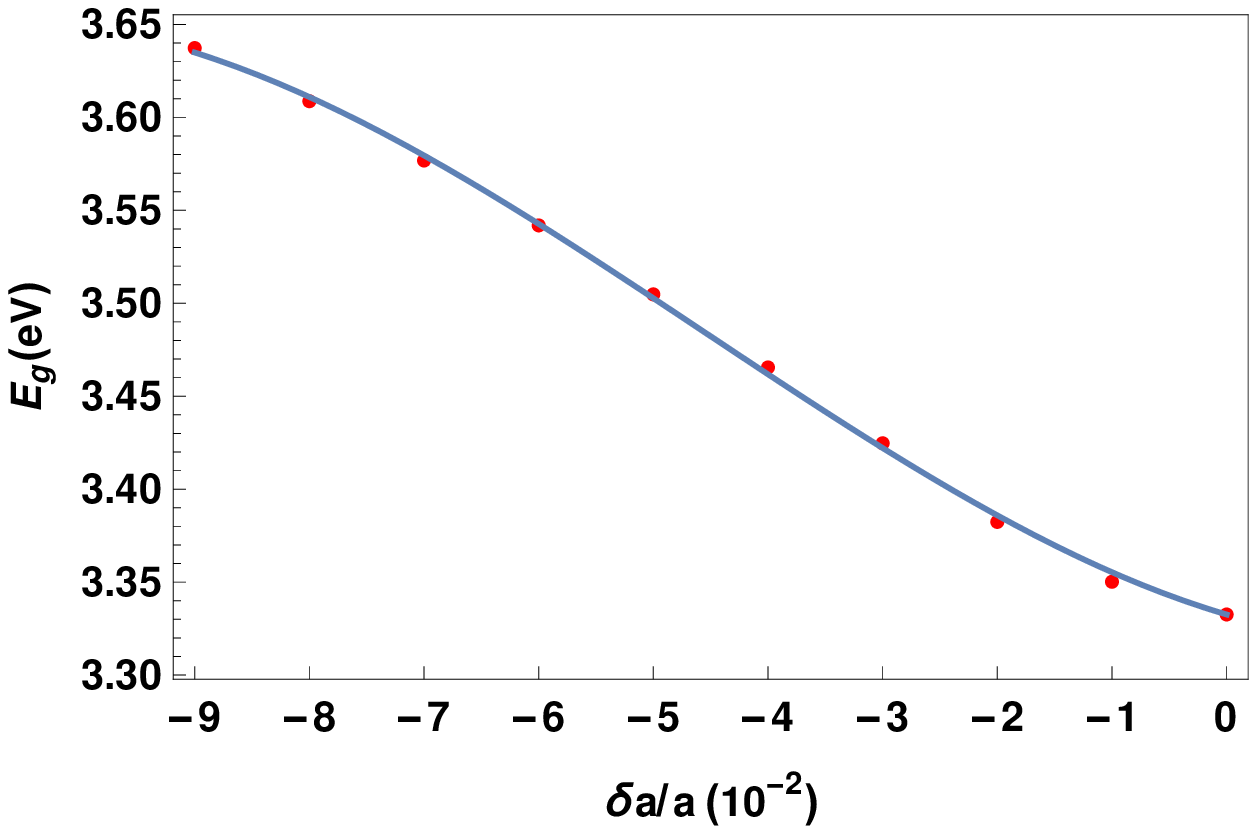}
	\caption{Calculated zone center energy band gap in zincblende GaN as a function of the biaxial strain deformation.}
	\label{FIG4}
\end{figurehere}

\noindent that this phenomenon also affects the energy positions of other conduction band minima ($L$, $X$), which could shift either upwards or downwards as a result of the reduction of the $xy$ lattice constant. However, since the energy distance between $\Gamma$ and the other minima is larger than $1\,$eV \cite{Bourgrov2001}, we hypothesize that the direct character of the fundamental gap does not alter.

Likewise, the top of the split-off band ($v3$), measured from the chosen zero energy reference, undergoes an increment as a result of further in-plane compressing, as shown in Fig. 5. This behavior, combined with the variation of the $v1$ band leads to the change of spin-orbit split-off energy $\Delta_{so}$ plotted in Fig. 6. This quantity 

\vspace{0.3cm}
\begin{figurehere}
	\centering
	\includegraphics[width=7.5cm]{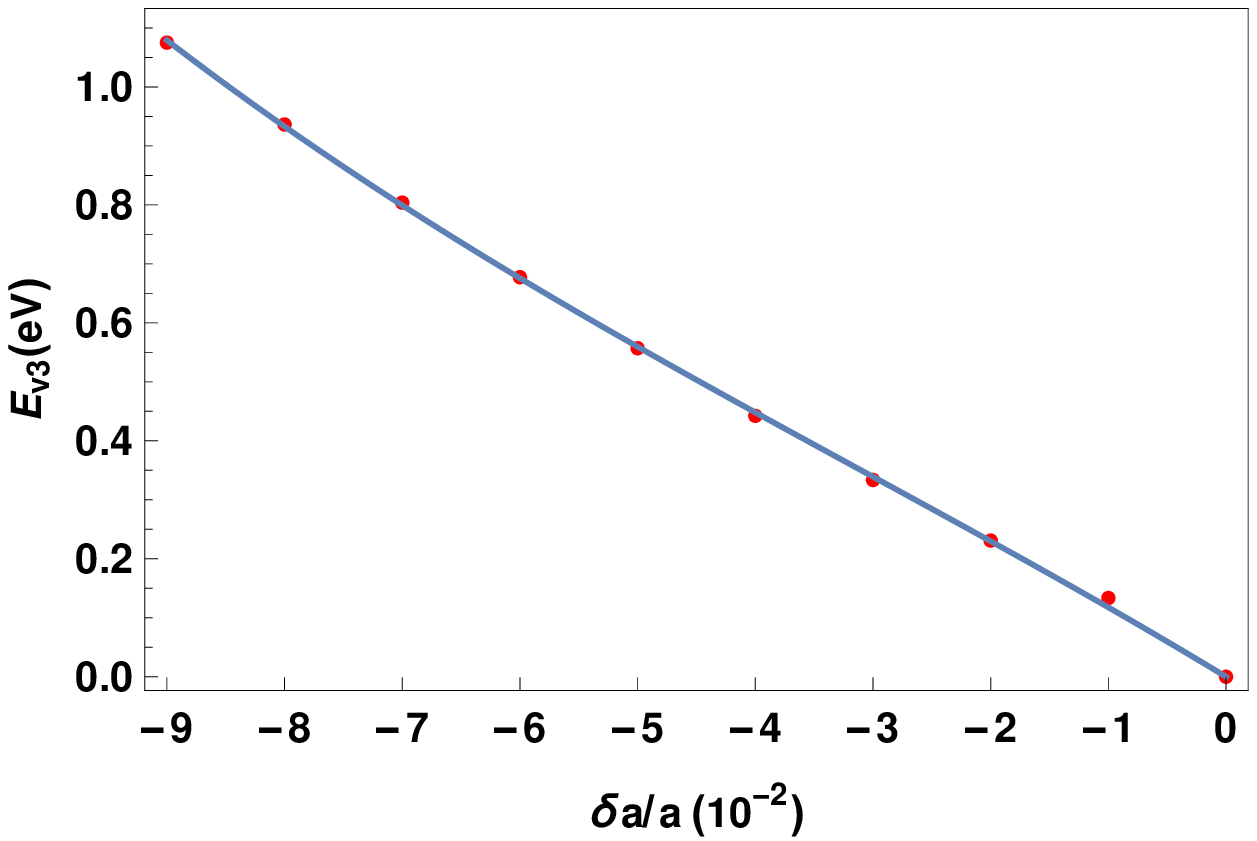}
	\caption{Calculated correction to the zone center energy of the split-off hole valence band ($v3$) top edge in zincblende GaN as a function of biaxial strain, measured from the unstrained value of the spin-orbit splitting energy ($\Delta_{so}(0)$).}
	\label{FIG5}
\end{figurehere}

\noindent shows a significant increase as a result of the biaxial deformation, showing an increment of approximately $1.2\,$eV above the unstrained value for sufficiently high compression.

The calculation yields a value $\Delta_{so}(0)=0.0251\,$eV for the unstrained ZB GaN. This is significantly larger than the accepted in the case of this material, which is of $0.017\,$eV \cite{Vurgaftman2003}. In another work, specifically devoted to evaluate it in III-V nitrides, Cardona and Christensen report $\Delta_{so}(0)=0.0185\,$eV \cite{Cardona2000}. More recently, the work by Bechstedt \textit{et al.}, on the first-principles calculation of the electronic structure of III-V polytypes, contains an 

\vspace{0.3cm}
\begin{figurehere}
	\centering
	\includegraphics[width=7.5cm]{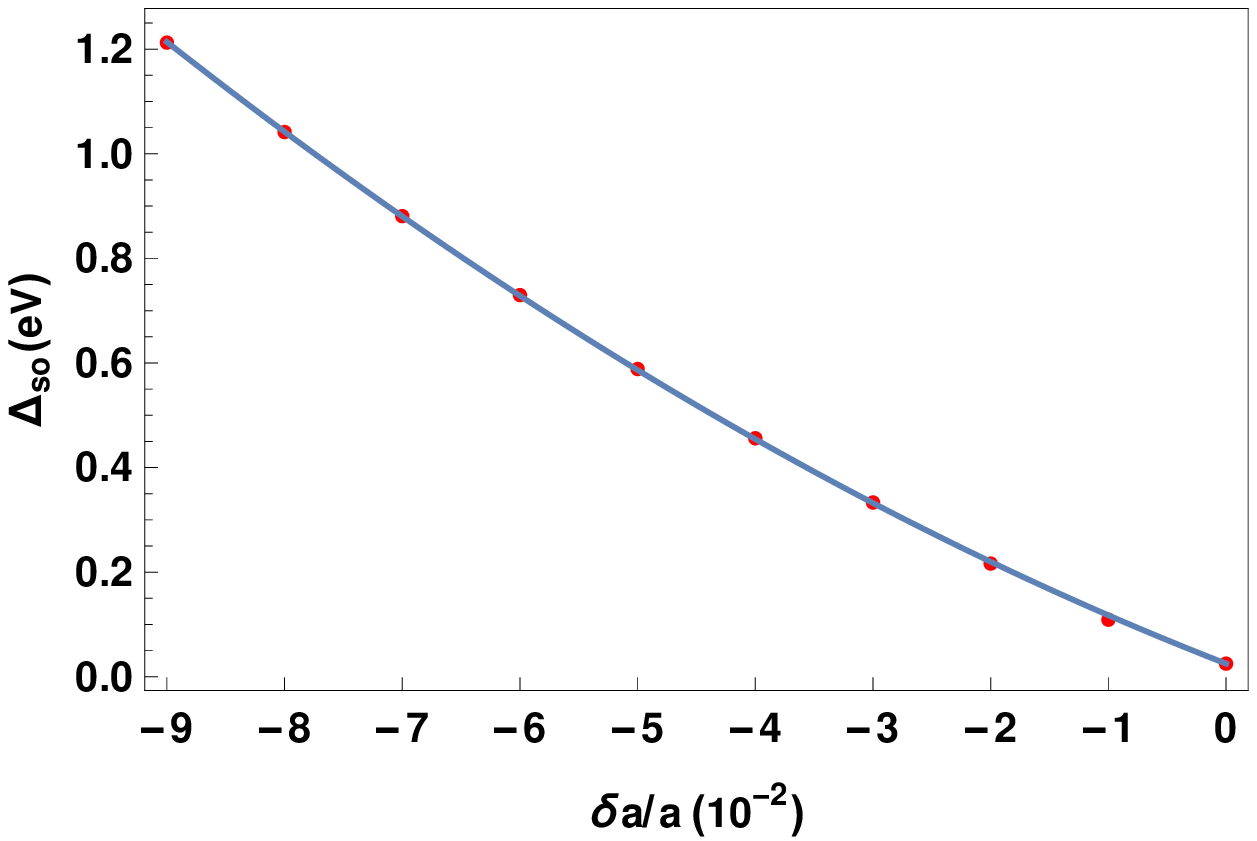}
	\caption{Calculated energy of valence band splitting due to spin-orbit interaction in zincblende GaN as a function of biaxial strain.}
	\label{FIG6}
\end{figurehere}

\noindent  even smaller value $\Delta_{so}(0)=0.015\,$eV \cite{Bechstedt2013}. This means that our calculation seems to overestimate the actual spin-orbit energy splitting in ZB GaN in a range of $7-10\,$meV approximately. Nonetheless, this energy keeps being noticeably small compared, for instance, with the energy band gap of the material, which is in agreement with all known results. Indeed, what is readily apparent is the striking increase of this quantity as a consequence of the lattice strain. This reflects in values above $1\,$eV for deformations close to $8\%$ and above.

%\subsection{Nombre de la subsecci\'{o}n}

\subsection{Biaxially strained ZB InN}

In an analogous way to that above commented for cubic GaN, here we are going to discuss the results of the DFT+HSE06(mod) calculation for the zone center electronic states of ZB InN subject to lattice biaxial compression.

In this context, Figs. 7-9 contain the evolution of energy position correction, measured from unstrained values, of the bottom of lowest conduction band ($E_c$) and tops of heavy- ($E_{v1}$) and light-hole ($E_{v2}$) valence bands, respectively. 

\vspace{0.3cm}
\begin{figurehere}
	\centering
	\includegraphics[width=7.5cm]{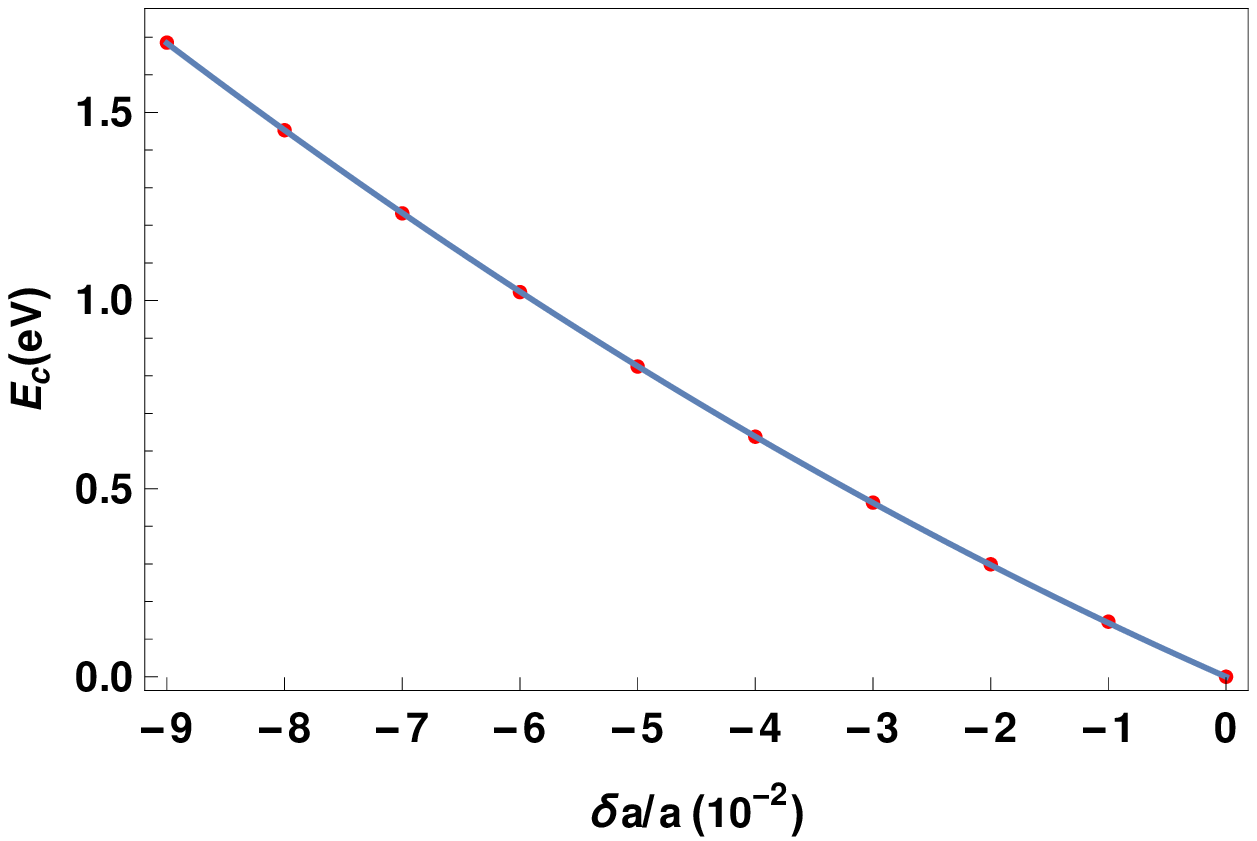}
	\caption{Calculated correction to the zone center energy of the conduction band bottom edge in zincblende InN as a function of biaxial strain, measured from the unstrained value.}
	\label{FIG7}
\end{figurehere}

Again, the quantity depicted, in each case, is the correction to the unstrained energy position (having chosen the zero of energies at the corresponding heavy-hole band one). It is possible to noticed the same kind of blueshift for energies already observed in ZB GaN.

\vspace{0.3cm}
\begin{figurehere}
	\centering
	\includegraphics[width=7.5cm]{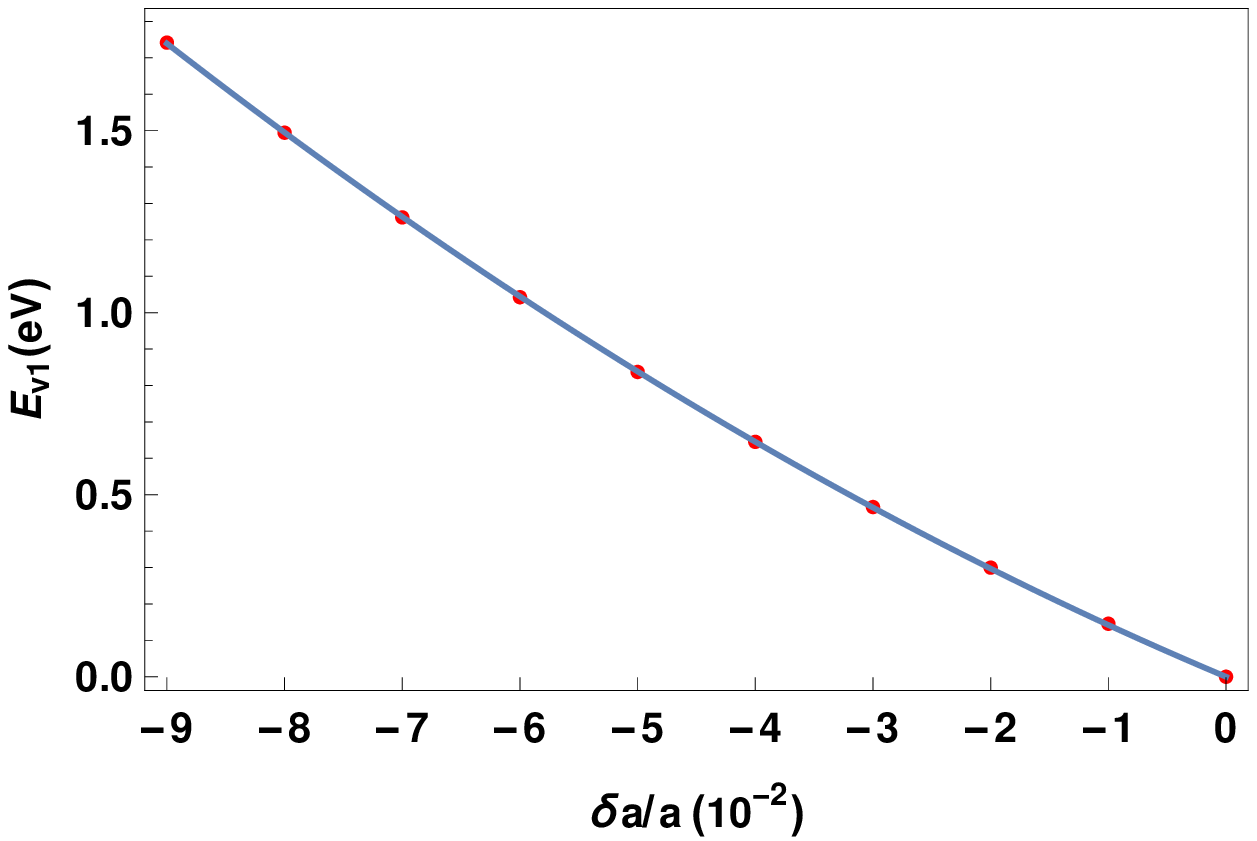}
	\caption{Calculated correction to the zone center energy of the heavy hole valence band ($v1$) top edge in zincblende InN as a function of biaxial strain, measured from the unstrained value, which is taken as the zero energy reference in this work.}
	\label{FIG8}
\end{figurehere}

\vspace{0.3cm}
\begin{figurehere}
	\centering
	\includegraphics[width=7.5cm]{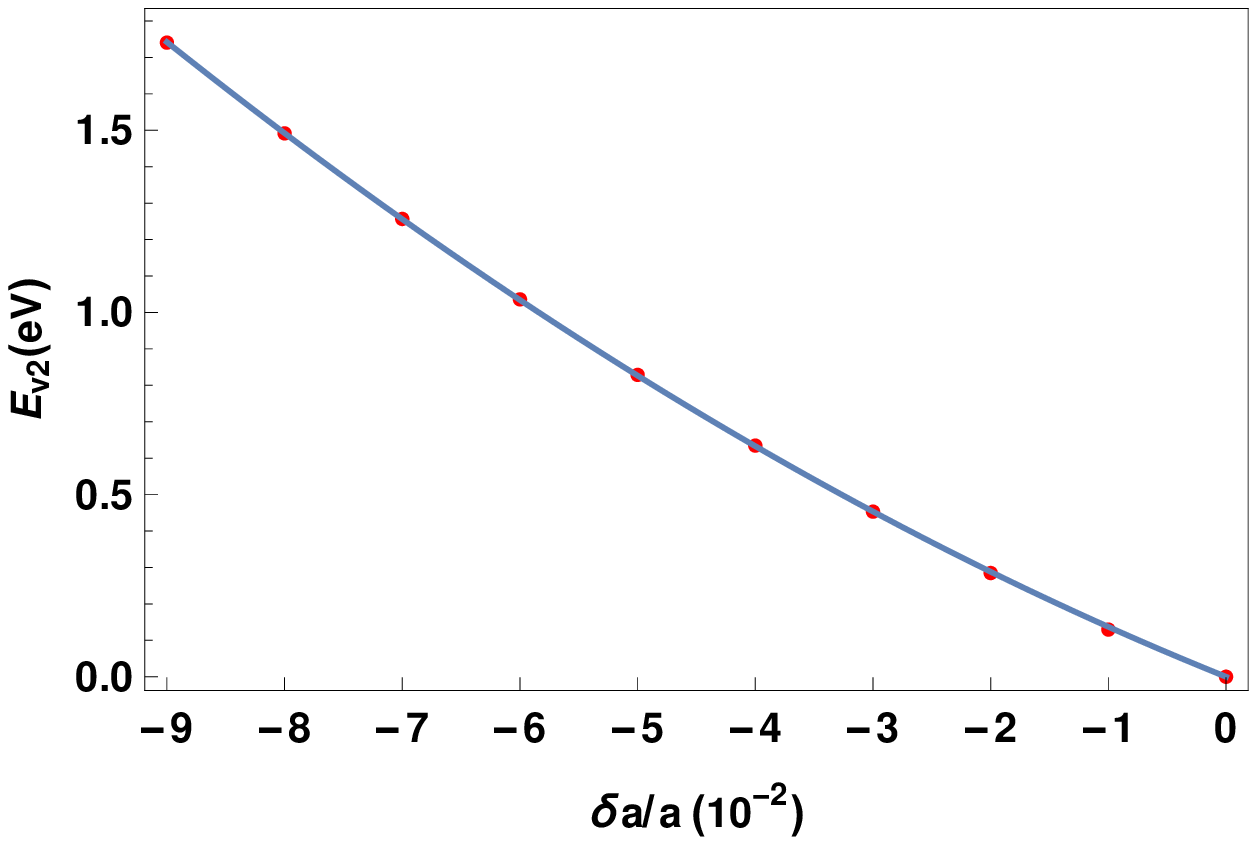}
	\caption{Calculated correction to the zone center energy of the light hole valence band ($v2$) top edge in zincblende InN as a function of biaxial strain, measured from the unstrained value.}
	\label{FIG9}
\end{figurehere}

In the Fig. 10 we are presenting the direct energy band gap of ZB InN as a function of the compressive biaxial strain, calculated from the corresponding variations of the conduction and heavy-hole band edges. It is worth recalling here that there is, still, certain degree of uncertainty with regard to the intrinsic value of $E_g$ in this material. Values going from $0.52\,$eV (considered as de the accepted one in \cite{Rinke2008}) and $0.533\,$eV (from \cite{Bechstedt2013}) to $0.78\,$eV (recommended by Vurgaftman and Meyer \cite{Vurgaftman2003}) can be found in the literature. Actually, there is consensus about the possible influence of Moss-Burstein effect since, in practice, a rather large density of donor atoms appear during the compound growth. Therefore, for our evaluation, we chose to reproduce the low temperature experimental result of $E_g(0)=0.61\,$eV reported by Sch\"ormann \textit{et al.} in high-purity c-InN \cite{As2006}. This process led to the already mentioned value $\alpha=0.40$ for the 

\vspace{0.3cm}
\begin{figurehere}
	\centering
	\includegraphics[width=7.5cm]{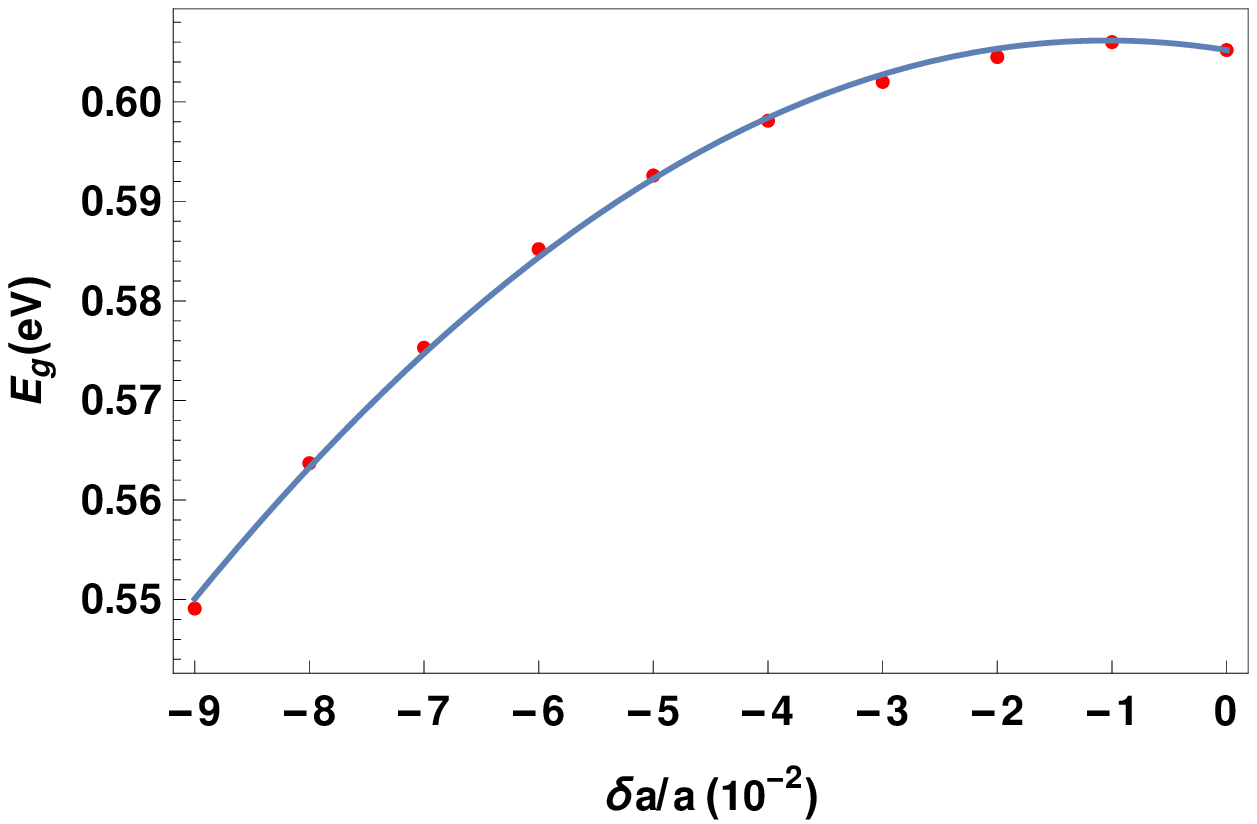}
	\caption{Calculated zone center energy band gap in zincblende InN as a function of the biaxial strain deformation.}
	\label{FIG10}
\end{figurehere}

\noindent DFT+HSE(mod) approach, and yields an unstrained value of $E_g=0.6052\,$eV, with a good agreement with the experimental one selected to reproduce.

In contrast to what happens with $E_g$ in c-GaN, after a very slight increment for small deformations, the gap of c-InN exhibits a drop that amounts $55\,$meV,  approximately, at the maximum $\vert \sigma\vert$ considered. This is equivalent to slightly $10\%$ below the relaxed crystal energy gap.

\vspace{0.3cm}
\begin{figurehere}
	\centering
	\includegraphics[width=7.5cm]{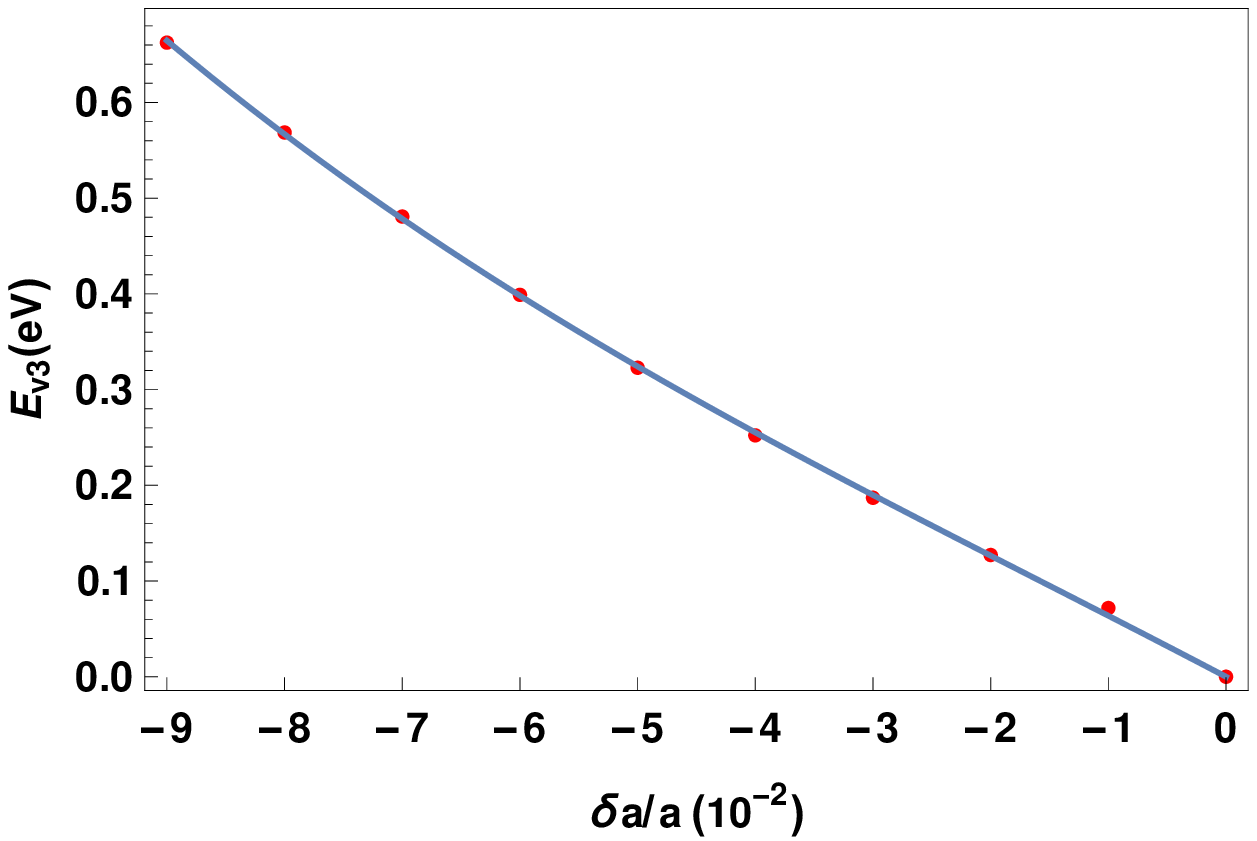}
	\caption{Calculated correction to the zone center energy of the split-off hole valence band ($v3$) top edge in zincblende InN as a function of biaxial strain, measured from the unstrained value of the spin-orbit splitting energy ($\Delta_{so}(0)$).}
	\label{FIG11}
\end{figurehere}

The upper edge of the spin-orbit split-off band also undergoes a blueshift as a result of the compressive biaxial deformation, as observed from Fig. 11. This also reflects in the notorious growth of the splitting energy $\Delta_{so}$, which goes from the unstrained value $\Delta_{so}(0)=0.032\,$eV to a value above $1\,$eV. In fact, our zero strain result turns out to be considerably grater than previously reported values such as $0.005\,$eV
\cite{Vurgaftman2003} and $0.0126\,$eV \cite{Cardona2000}. However, it is only $11\,$meV above the much more recent result by Bechstedt and collaborators, who obtained $\Delta_{so}(0)=0.021\,$eV \cite{Bechstedt2013}. Notice that the later report became the first one in which this quantity appears to be larger in ZB InN, compared with its corresponding one in ZB GaN. Our result, in this sense, coincides with the work in \cite{Bechstedt2013}.

\vspace{0.3cm}
\begin{figurehere}
	\centering
	\includegraphics[width=7.5cm]{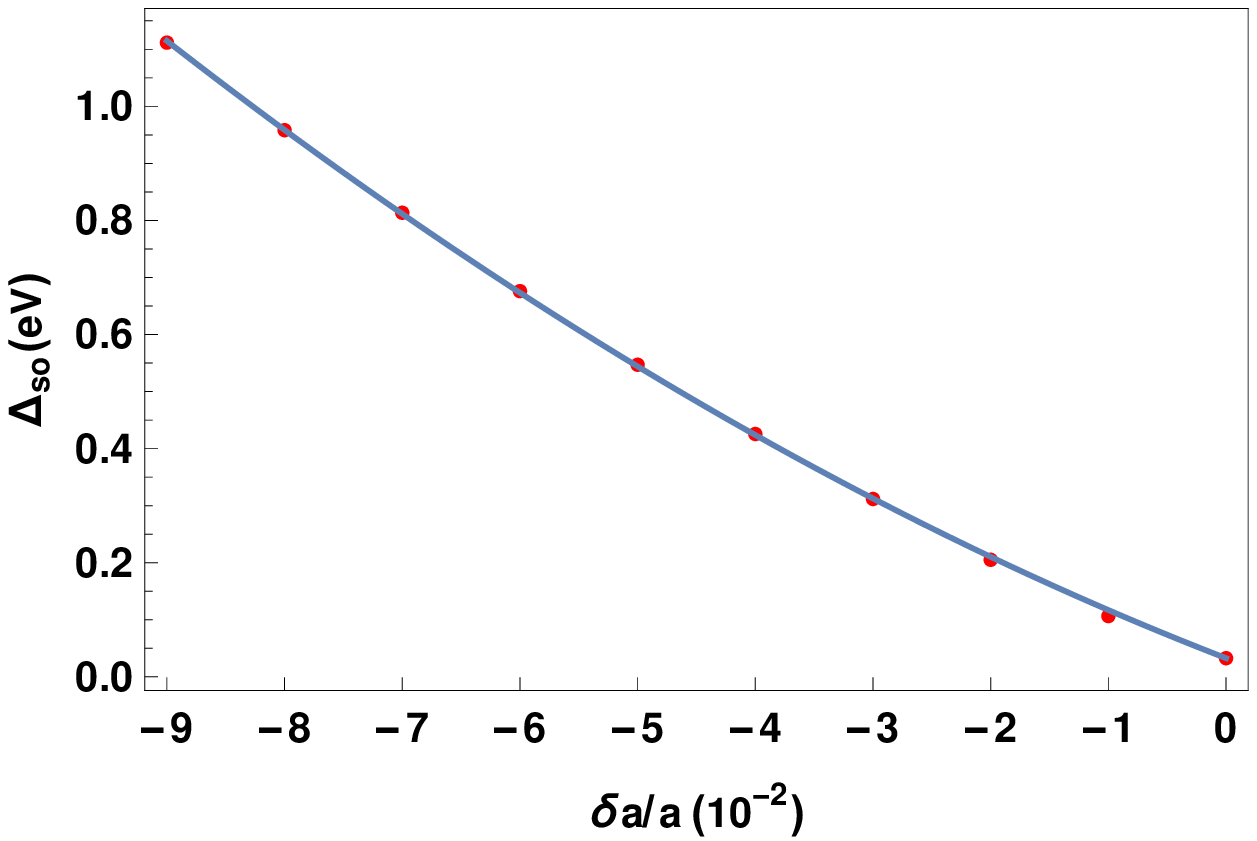}
	\caption{Calculated energy of valence band splitting due to spin-orbit interaction in zincblende InN as a function of biaxial strain.}
	\label{FIG12}
\end{figurehere}

\subsection{Effect of hydrostatic deformation on the zone center band structure in zincblende GaN}

Since hydrostatic deformation does not alter the cubic geometry of the elementary cell, the $\Gamma$-point degeneration 

\vspace{0.3cm}
\begin{figurehere}
	\centering
	\includegraphics[width=7.5cm]{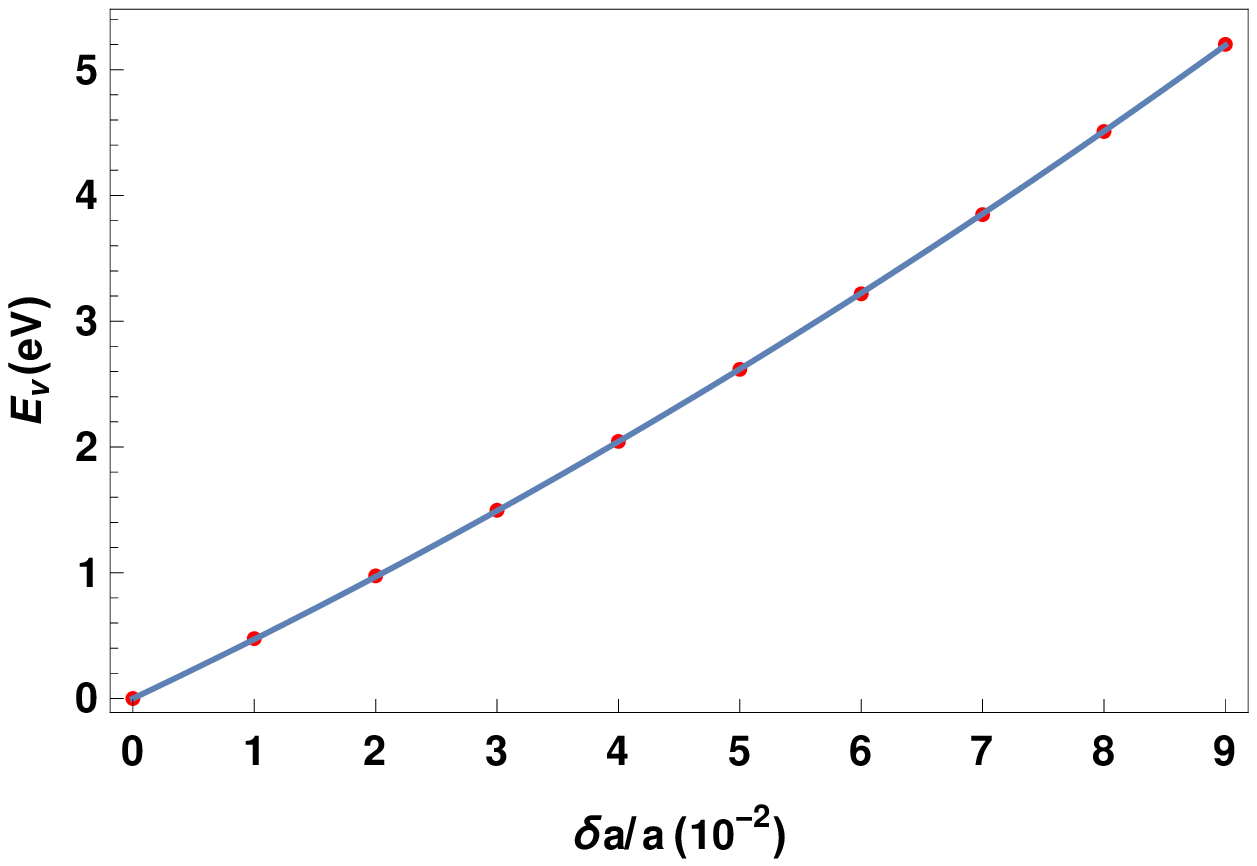}
	\caption{Calculated variation of the upper edge of the valence band in zincblende GaN, as a function of hydrostatic deformation, measured from the undeformed result (taken as the zero for energies).}
	\label{FIG13}
\end{figurehere}

\noindent of the upper two valence bands (heavy-hole and light-hole) is preserved. Therefore, our report for $E_v(\epsilon)$, corresponds to both heavy- and light-hole edges $E_{v1}$ and $E_{v2}$, as we have named them above. They shift together due to the compression. For the same reason, we are not going to present the variation of the upper edge of the spin-orbit slit-off band for it coincides with that of $\Delta_{so}$. In this sense, Figs. 13 to 16 contain, in that order, the variation of valence band top edge energy position (taken to be the origin of energies in the situation of zero deformation), the correction to conduction band bottom position (measured from its $\epsilon=0$ value), the zone center energy band gap as a function of hydrostatic compression, and the functional dependence $\Delta_{so}(\epsilon)$.

\vspace{0.3cm}
\begin{figurehere}
	\centering
	\includegraphics[width=7.5cm]{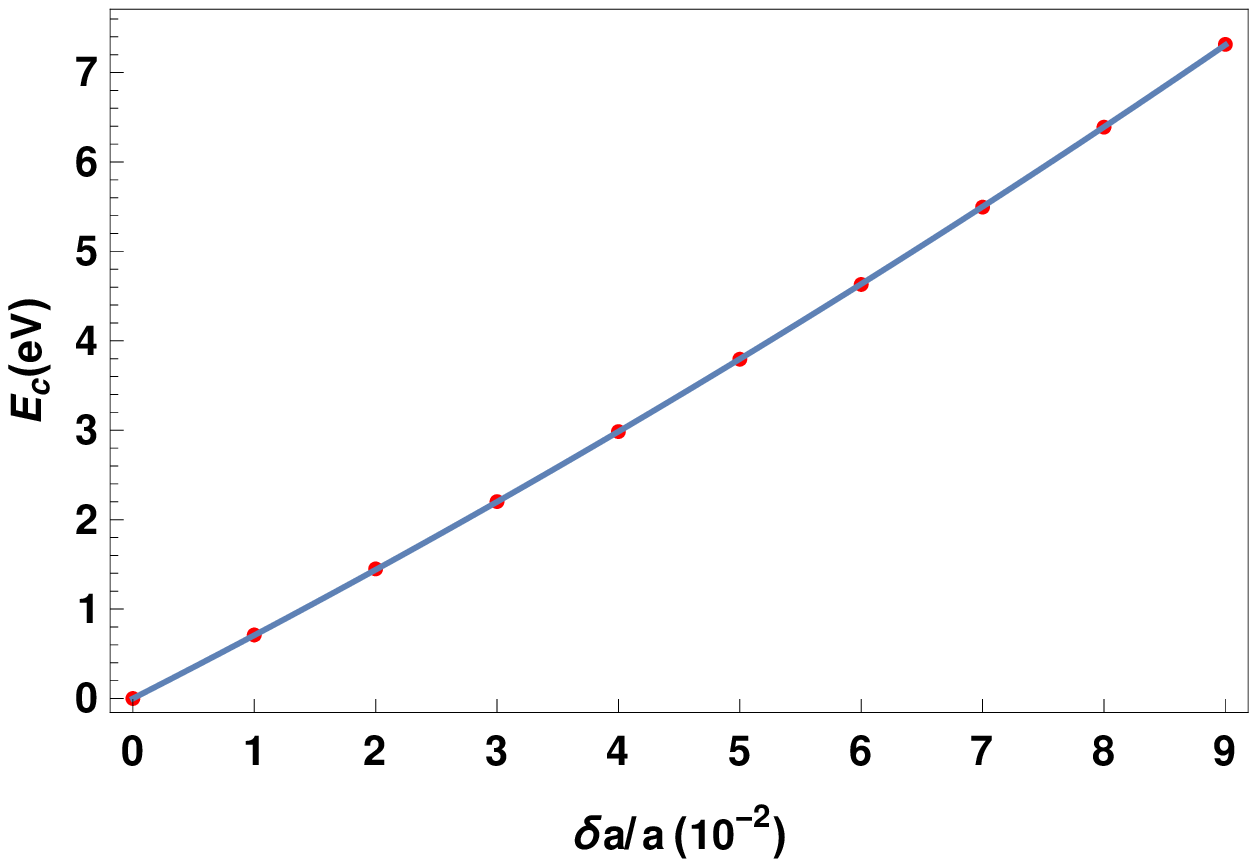}
	\caption{Calculated variation of the zone center conduction band bottom in zincblende GaN, as a function of hydrostatic deformation, measured from the undeformed result.}
	\label{FIG14}
\end{figurehere}

\vspace{0.3cm}
\begin{figurehere}
	\centering
	\includegraphics[width=7.5cm]{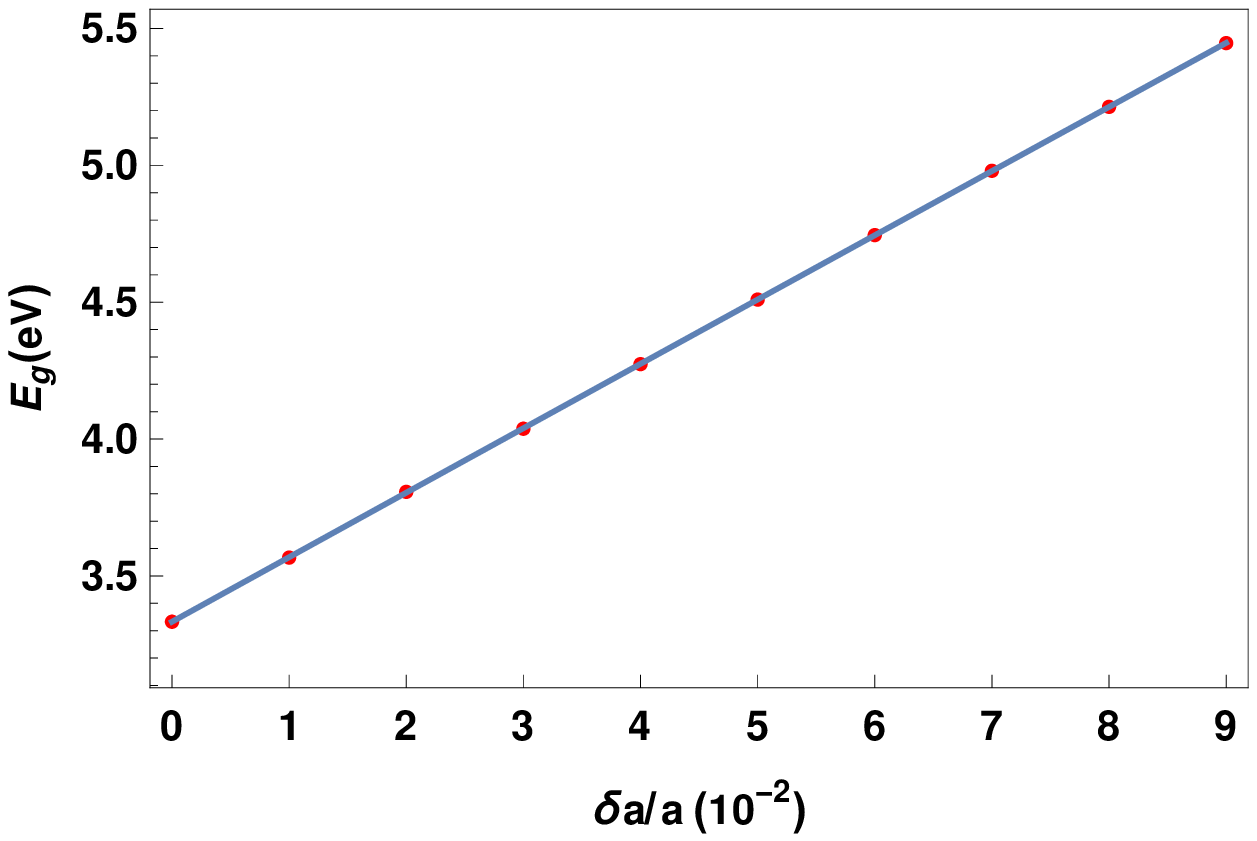}
	\caption{The variation of the direct energy band gap of zincblende GaN as a function of hydrostatic deformation}
	\label{FIG15}
\end{figurehere}

From the results depicted one may observe that, once again, the hydrostatic compression causes the displacement towards higher energies of the band edges as well as of the direct band gap. The latter exhibits a significant increase of $2\,$eV, approximately. This is equivalent to $1.6$ times the value of $E_g$ without deformation when the reduction in the side of the elementary cell reaches a $91\%$ of the undeformed one. This is contrast to the increment by about a $10\%$ of the band gap above reported in the case of a biaxial compression, thus confirming the rather different natures of both kinds of deformations. 

Besides this, it turns out that hydrostatic compression also acts in a very different way in what respects to its effect on $\Delta_{so}$. As shown in Fig. 16, this quantity has a decreasing behavior and the amount of its change is much less pronounced that those obtained in the case of biaxial compression, with a tendency to reduction on which we shall more extensively comment below in the case of InN.

\vspace{0.3cm}
\begin{figurehere}
	\centering
	\includegraphics[width=7.5cm]{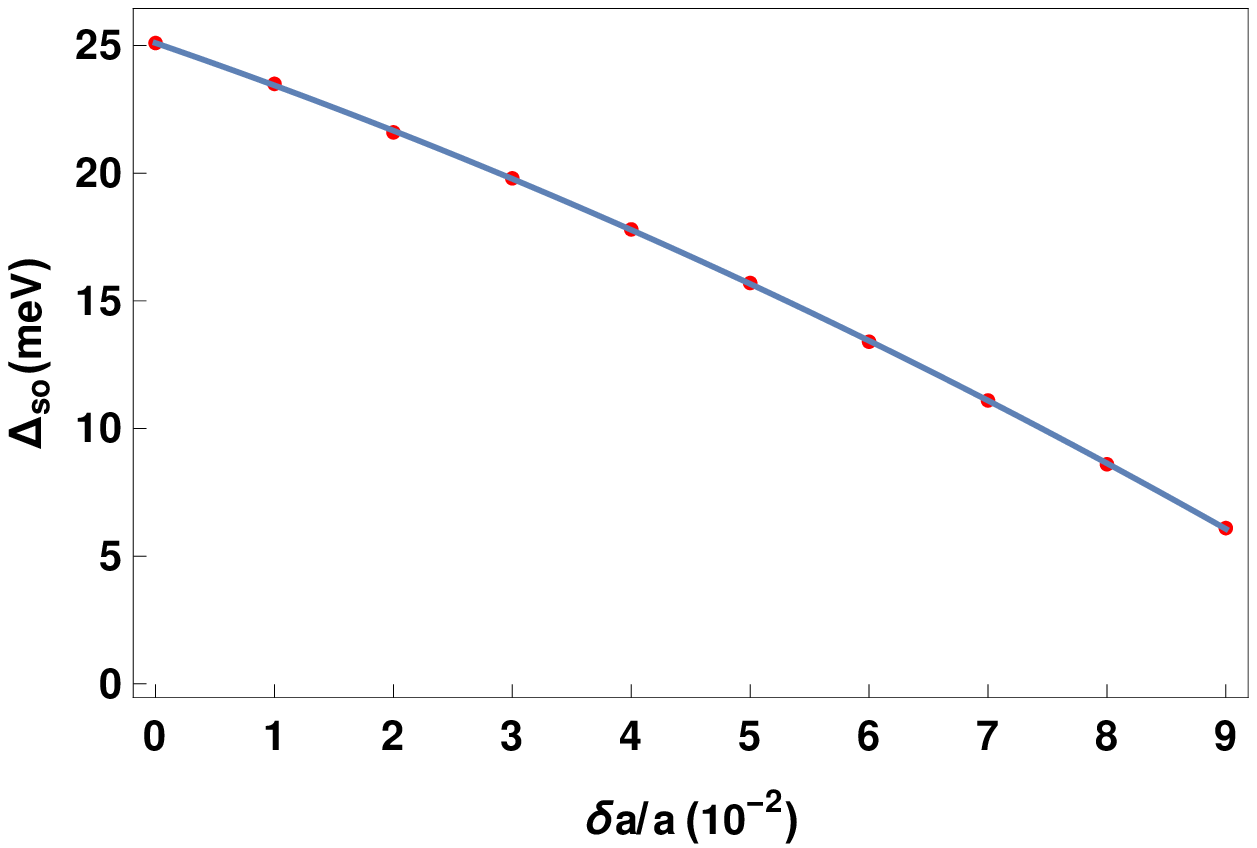}
	\caption{Calculated spin-orbit splitting energy in zincblende GaN as a function of the magnitude of the hydrostatic deformation.}
	\label{FIG16}
\end{figurehere}

\subsection{Effect of hydrostatic deformation on the zone center band structure in zincblende InN}

In regard to this material, c-InN, Figs. 17-20 contain, respectively, the variation with respect to the magnitude of hydrostatic deformation of the  upper valence band edge energy position ($E_v$), the $\Gamma$-point conduction band boot, ($E_c$), the width of the direct energy gap ($E_g$), and the split-off valence band energy ($\Delta_{so}$). 

\vspace{0.3cm}
\begin{figurehere}
	\centering
	\includegraphics[width=7.5cm]{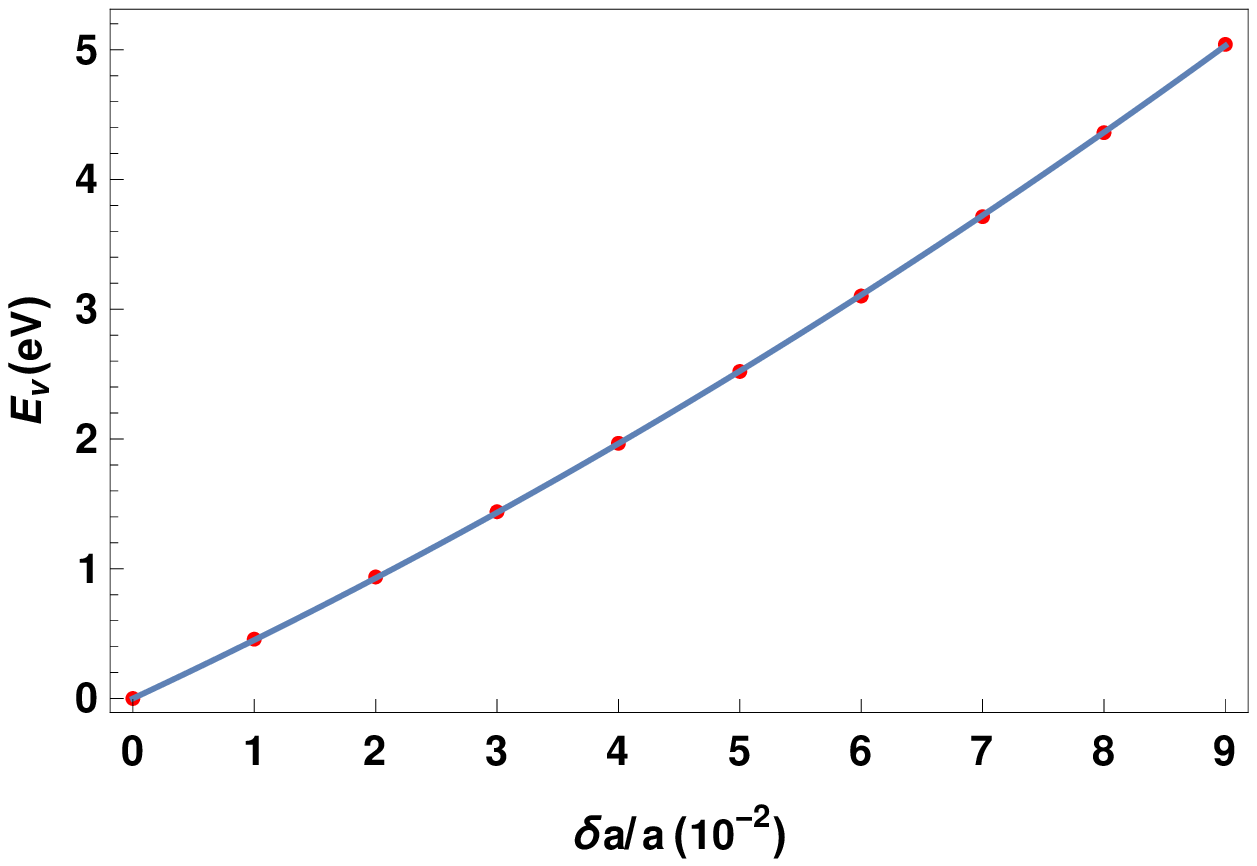}
	\caption{Calculated variation of the upper edge of the valence band in zincblende InN, as a function of hydrostatic deformation, measured from the undeformed result (taken as the zero for energies).}
	\label{FIG17}
\end{figurehere}

\vspace{0.3cm}
\begin{figurehere}
	\centering
	\includegraphics[width=7.5cm]{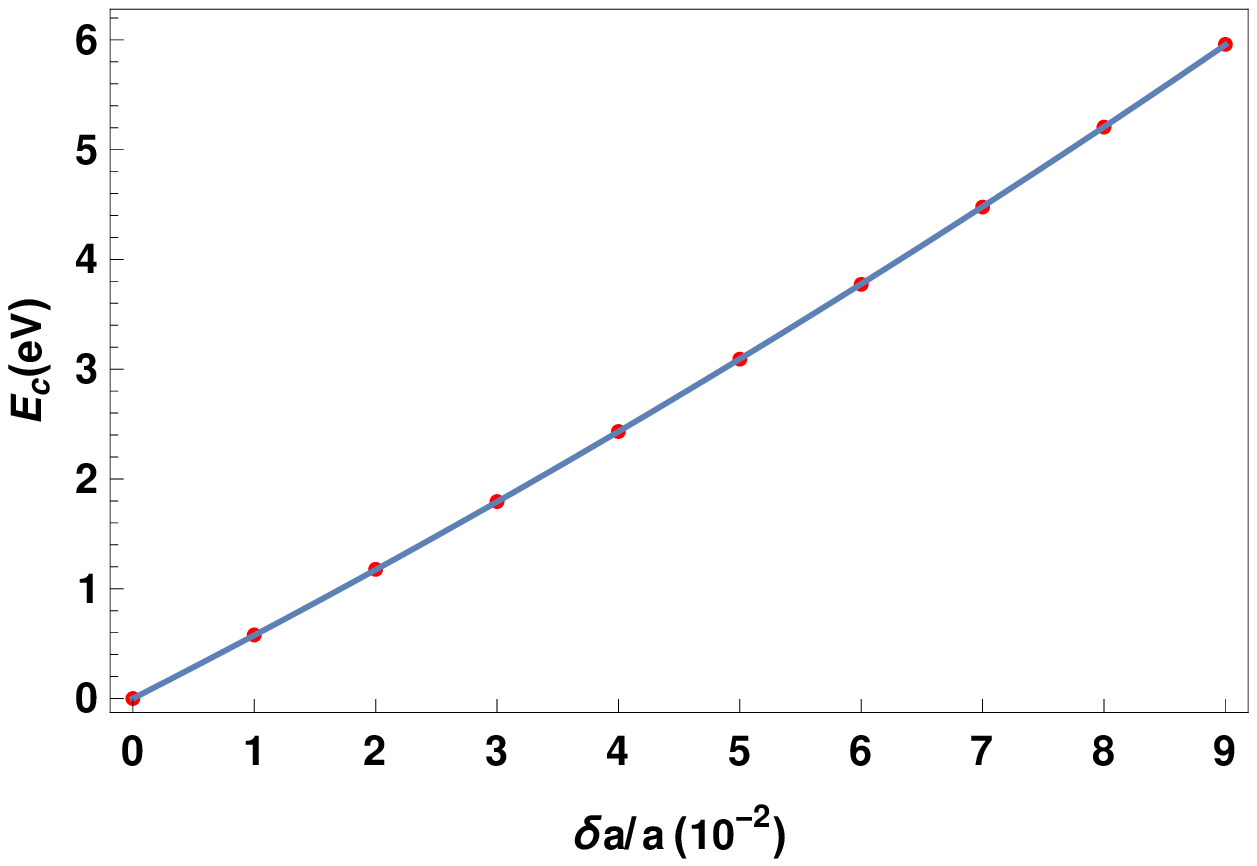}
	\caption{Calculated variation of the zone center conduction band bottom in zincblende InN, as a function of hydrostatic deformation, measured from the undeformed result.}
	\label{FIG18}
\end{figurehere}

Once more, we find that $k=0$ band edges undergo a shift to the blue, together with the widening of the forbidden energy gap as a results of the hydrostatic deformation. The change in $E_g$, in this case, implies an increase of about two times with respect to the gap of the uncompressed material, and also noticeably contrasts with the slight diminishing associated to the effect of biaxial compression.

\vspace{0.3cm}
\begin{figurehere}
	\centering
	\includegraphics[width=7.5cm]{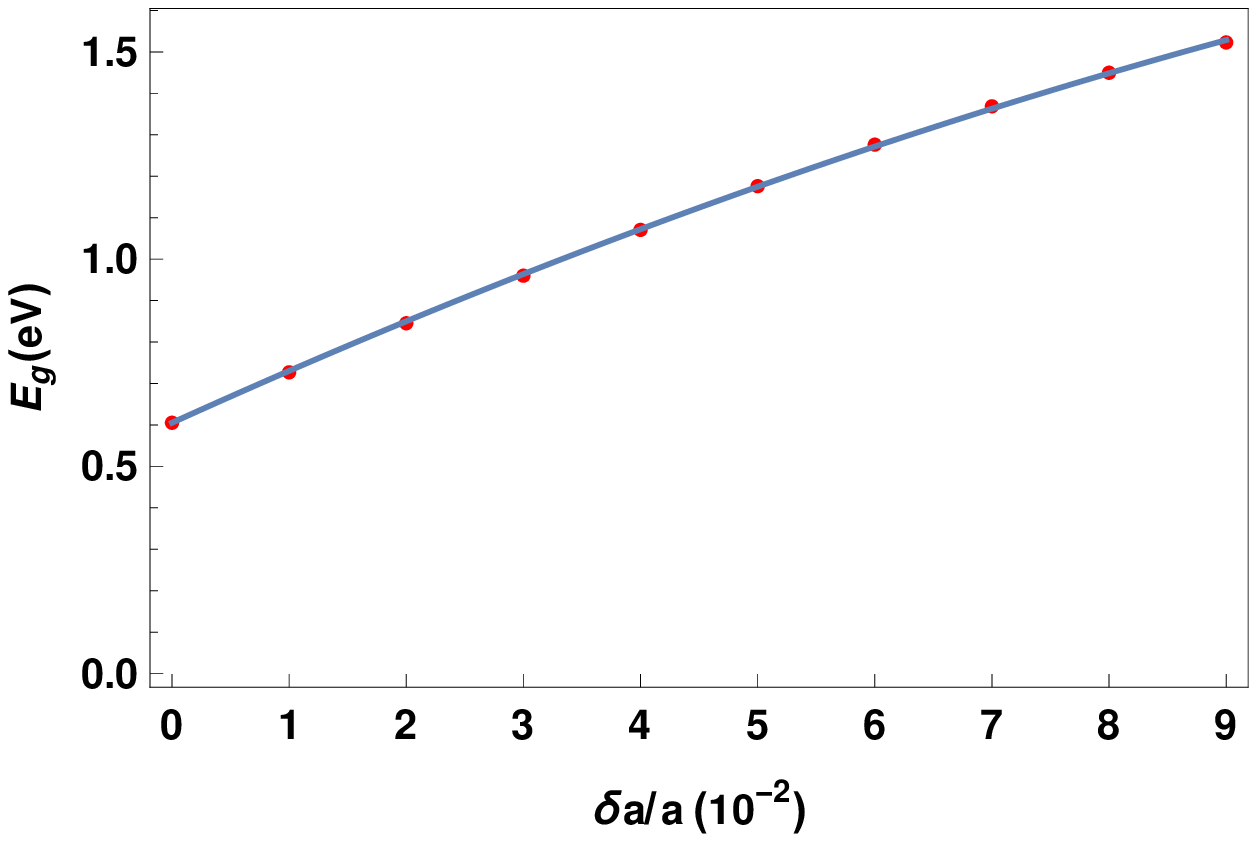}
	\caption{The variation of the direct energy band gap of zincblende InN as a function of hydrostatic deformation}
	\label{FIG19}
\end{figurehere}

Special mention requires the phenomenon of variation in the split-off energy induced by hydrostatic deformation. As the reader may observe from Fig. 20(a), the fitting curve of $\Delta_{so}$ extends only until a $7\%$ of lattice constant reduction. For higher values of compression, the decreasing trend of this quantity reverses.

\vspace{0.3cm}
\begin{figurehere}
	\centering
	\includegraphics[width=7.5cm]{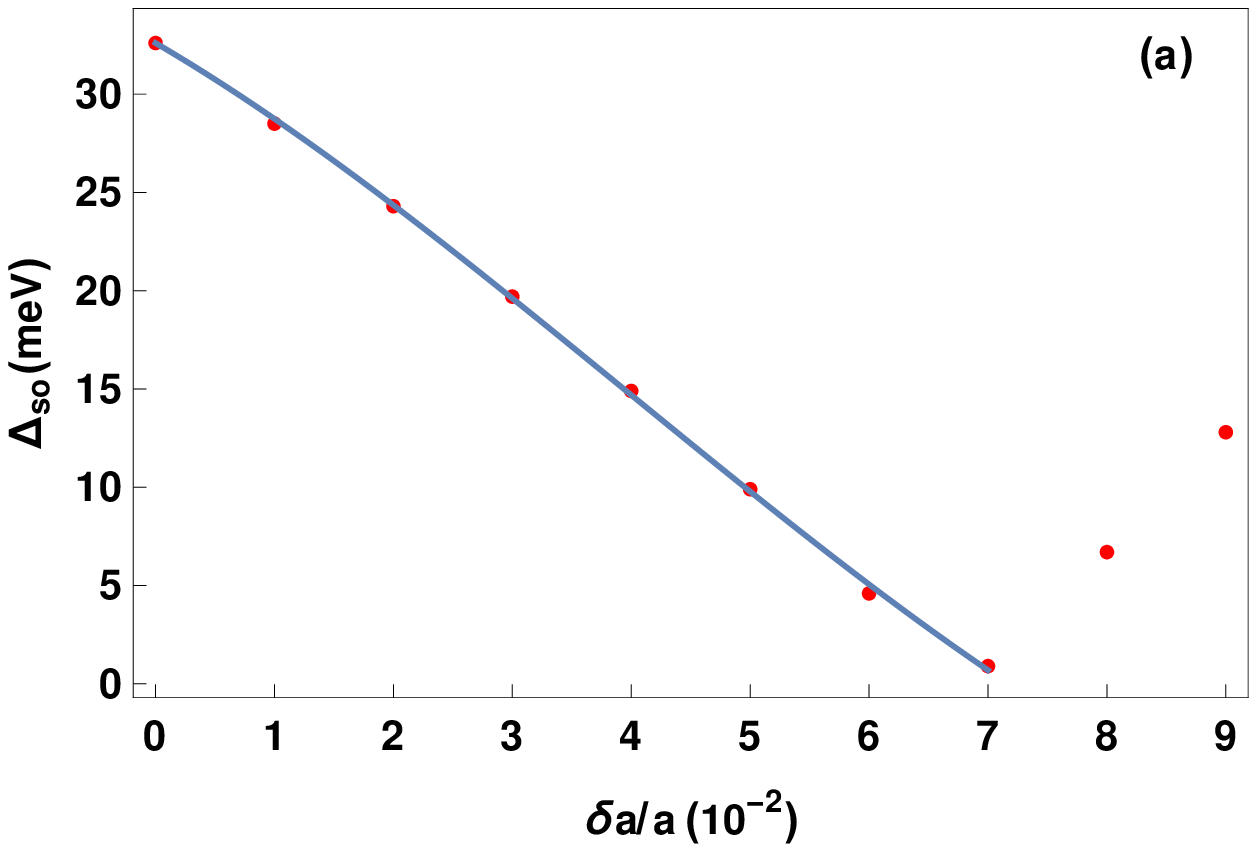}
	\includegraphics[width=7.5cm]{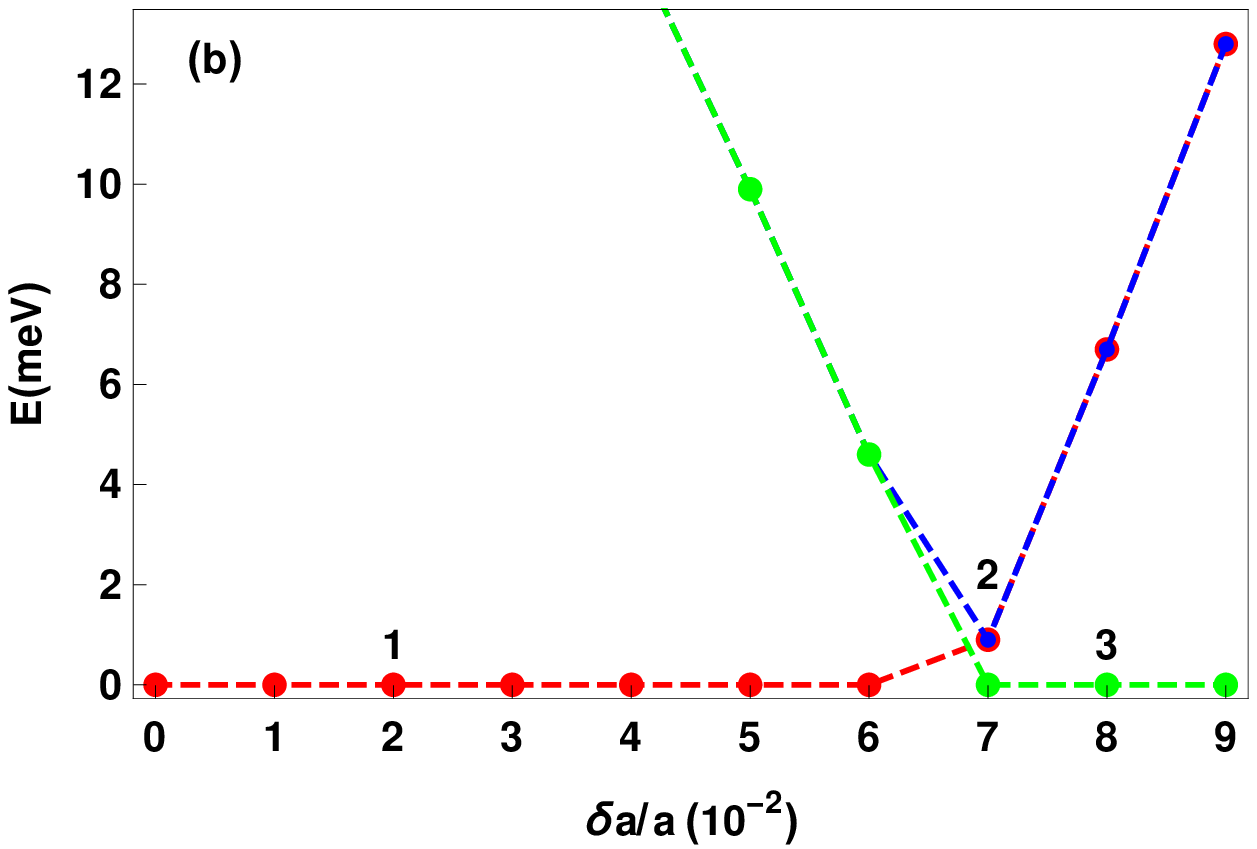}
	\caption{(a) Energy of spin-orbit valence band splitting in zincblende InN as a function of the magnitude of hydrostatic deformation. (b) Comparison of the second (light-hole) and third (split-off) valence band edge variations.}
	\label{FIG20}
\end{figurehere}

It is worth discussing what happens with the displacement of valence bands edges. Accordingly, Fig. 20(b) contains the zone center energy distance of the second and third valence bands (lines 1 and 3 in the figure), with respect to the the energy position of the first (top) one as a function of hydrostatic deformation magnitude. One observes that, until a value $\epsilon=0.06$, the top edges of light and heavy hole bands coincide ($E_{v1}$) and, clearly, the energy difference between them is zero (line 1) Meanwhile, the separation between the split-off band and $E_{v2}$ is reducing and, for $\epsilon=0.07$, both bands have merged and keeps evolving together at the time that they start to distance themselves from $E_{v1}$ (line 3. Remind that this position is the zero for energy difference calculation). The change of energy position corresponding to this joint edge is the one depicted by isolated points (not joined by a solid line) in Fig. 20(a).

Our proposal to explaining this phenomenon lies on the presence of a structural phase transition, induced by hydrostatic compression (see, for instance, Refs. \cite{Daoud2018, Kunc2015,Schwarz2014,Saoud2011,Verma2010,Duan2010,Cheng2007,Silva2005,Serrano2000, Pandey1993,Gorczyca1993}). It is known that this transition may appear both for WZ and ZB III-V nitrides.It implies a change in the crystal structure to that of rocksalt (RS), which belongs to the cubic system ($fcc$ with four cations and four anions per cell). 

Given that our calculation does not include the structural analysis of the material, we have relied upon analogous situations, reported in the literature. In particular, the work \cite{Segura2003} shows the distribution of valence bands in pressure-induced RS ZnO. The phase transition to RS structure seems to imply an indirect gap configuration, as indicated in the case of ZnO and as shown in the recent article by Kumar and Roy, where the authors investigate six different crystalline phases of InN via first-principles DFT  \cite{Kumar2018}. Precisely, in the case of RS, they report the main gap to occur between the $L$ and $\Gamma$ points. Noticeably, the calculation environment used in that work is, also, \textit{Quantum Espresso}; but it uses a GGA+PBE scheme that leads to zero direct gap in the WZ and ZB cases. This clearly contradicts the experiments and, therefore, serves as an additional reason to justify the need of using a modified HSE approach as the one presented here.

Then, by virtue of the above comments, we can consider that the effect of hydrostatic compression on the Brillouin zone center energy states of ZB InN would be correctly described until a deformation magnitude of about $6-7\%$ in the lattice parameter. Values above it will not be describing the behavior of those states in the compound of our interest. However, our investigation suggests that, for the ZB GaN case, the results would be adequate within the entire range of deformations considered.
 
\subsection{The conduction band effective mass and other quantities}

We have left to separately discuss the matter of the calculation of conduction band bottom effective mass, $m^*/m_o$, in ZB GaN and InN. A technical inconvenient appeared when trying to draw the energy bands for $k$ large enough within the Brillouin zone using the chosen  computational packages. The main problem arises at the time of including spin-orbit corrections, with the particular incidence of the smallness in $\Delta_{so}$, typical of these compounds. The difficulty is more pronounced in the case of biaxial deformation given the related break of the original cubic symmetry in the unit cell. 

In consequence, we had to limit ourselves to analyzing the energy dispersion relation in the close vicinity of the conduction band bottom,  along the $\Gamma-X$ direction, and extend the deformation only up to a $7\%$. Something that could be done in straightforward manner was to determine the corresponding values of under zero deformation conditions. This lead to the a result of $m^*/m_o=0.1920$ for ZB GaN and to $m^*/m_o=0.0500$ for ZB InN. These values are very close to the $0.1930$ and $0.054$, respectively accepted for these materials as reported in the work of Rinke \textit{et al.} \cite{Rinke2008}. This coincidence validates, as well, the pertinence of modifying the DFT+HSE scheme. Although the change in $\alpha$ mainly orients to correctly reproduce the experimental energy band gap, the fact of achieving good agreement for conduction effective mass, too, reinforces its validity.

Table I contains the calculated $m^*/m_o$ values as function of both the compressive biaxial strain and the hydrostatic deformation. In general, the effective mass shows and initial increase as a result of diminishing the lattice constant. But, then, above certain magnitude of $\delta a/a$, in each case, this quantity decreases. This is something already seen in the tight-binding analysis of strain effect in ZB GaAs and InAs, reported in \cite{Mozo2021}. In the hydrostatic case, the tendency of $m^*/m_o$ is very similar for the two compounds, with a greater regularity in the variation what make us think that the computing tools have less problems when the cubic cell symmetry is preserved. 

Going over to the possible structural phase transition of InN under hydrostatic compression, we pay particular attention to the change in monotony shown by $m^*/m_o$ when $\epsilon$ goes above $0.06$, which could be regarded as the limit for keeping the ZB geometry. In fact, we were able to extend the calculation, in this case, up to $\epsilon=0.09$. What came out of this that $m^*/m_o=0.0993,\;0.1291,\;{\rm and}\;0.1505$, for $\epsilon=0.07,\;0.08,\;{\rm and}\;0.09$, respectively. That is, a complete alteration of the variation tendency, compared to that when the cell geometry is supposed to keep its original symmetry. Besides, the obtained increment is significantly pronounced. In our opinion, this could be an additional argument in favor of the hypothesis of a structural phase transition, induced by hydrostatic compression.

Finally, there are a couple of parameters of wide interest for which one could derive their corresponding variations as functions of $\sigma$ and $\epsilon$ in the materials under study. They are the conduction band nonparabolicity parameter, $\beta$, and the Kane energy, $E_P=2m_oP^2/\hbar^2$, where $P=-i(\hbar/m_o)\langle s\vert \hat{p}_z\vert z\rangle$ is a real quantity proportional to the matrix element of momentum operator between conduction and valence states. It is the unknown parameter in Kane's $\vec{k}\cdot\vec{p}$ model \cite{Kane1957} which needs to be determined either experimentally or by microscopic calculations. In the case of $\beta$, it can be determined from the expression derived by Vrehen within the $\vec{k}\cdot\vec{p}$ description  \cite{Vrehen1968} 

\[ \beta=\left (1-\frac{m^*}{m_o} \right) ^2\frac{3E_g+4\Delta_{so}+2\Delta_{so}^2/E_g}{(E_g+\Delta_{so})(3E_g+\Delta_{so})},\]

\noindent whilst the value of $E_P$ will come from the Kane formula 

\[\frac{m_o}{m^*}=1+\frac{E_g+\frac{2}{3}\Delta_{so}}{E_g(E_g+\Delta_{so})}E_P. \]

The dependence of these two parameters on the quantities calculated in the present work, makes possible to evaluate them, as functions of deformations, in an indirect way. 

\begin{widetext}
	
		\begin{table}[]
		\caption{Zincblende GaN and InN consuction band effective masses as functions of the magnitude of compressive biaxial and hydrostatic deformations.}
		\begin{center}
			\begin{tabular}{|c|c|c||c|c|}
				\hline
				\multicolumn{1}{|l|}{$\vert \delta a/a\vert$} & \multicolumn{2}{c||}{Biaxial compression} & \multicolumn{2}{c|}{Hydrostatic deformation}                         \\ \hline
				\multicolumn{1}{|l|}{} & \multicolumn{1}{l|}{GaN$\,(m^*/m_o)$} & \multicolumn{1}{l||}{InN$\,(m^*/m_o)$} & \multicolumn{1}{l|}{GaN$\,(m^*/m_o)$} & \multicolumn{1}{l|}{InN$\,(m^*/m_o)$} \\ \hline
				$0$ 	&    $0.1920$    &    $0.0500$    &     $0.1920$     &     $0.0500$   \\
				$0.01$	&    $0.1970$    &    $0.0510$    &     $0.1948$     &     $0.0569$   \\
				$0.02$	&    $0.1971$    &    $0.0580$    &     $0.2014$     &     $0.0679$   \\
				$0.03$	&    $0.1947$    &    $0.0671$    &     $0.2141$     &     $0.0750$   \\
				$0.04$	&    $0.1917$    &    $0.0664$    &     $0.2312$     &     $0.0950$   \\
				$0.05$	&    $0.1912$    &    $0.0588$    &     $0.2468$     &     $0.1044$   \\
				$0.06$	&    $0.1910$    &    $0.0513$    &     $0.2498$     &     $0.0944$   \\
				$0.07$	&    $0.1890$    &                &     $0.2475$     &                \\\hline
			\end{tabular}
		\end{center}
	\end{table}
\end{widetext}

\section{Conclusions}

We have performed a first-principles Hybrid Functional DFT investigation on the effect of compressive biaxial strain and hydrostatic deformation on the parameters that characterize the electron states at the Brillouin zone center in the III-V semiconductors gallium arsenide and indium arsenide, in the case when these compounds crystallize in the zincblende structure.

The calculation includes a search for a suitable value of Fock exchange percentage through the so-called "$\alpha$" parameter in the Heyd-Scuseria-Ernzerhof hybrid functional formalism, in order to correctly determine the accepted values of the direct energy band gaps in those materials. This value turns out to be different in each case, being equal to $0.43$ in the case of zincblende GaN and $0.40$ for zincblende InN. It is worth saying that, with the use of such $\alpha$-values, the conduction effective of the compounds, under zero deformation conditions, were correctly reproduced as well.

The compressive biaxial strain causes a general blueshift of the zone center conduction and valence band edges, with slight to moderate increases of the direct energy band gaps. Significantly high increments were obtained for the spin-orbit split-off valence band energies, with the obtained unstrained values quite above the generally accepted ones. This is, perhaps, the only negative aspect of our process, although one should keep in mind that precise experimental values of $\Delta_{so}$ do not seem to be available for cubic nitrides. Conduction effective masses show an initial increase with strain but after certain value of deformation they are decreasing functions of the biaxial compressive effect.

According to our results, hydrostatic deformation leads to a strong increasing of zincblende GaN energy gap, whereas in the InN it causes a slight overall reduction. However, for values of the magnitude of hydrostatic compression above $6\%$, InN seems to undergo a structural phase transition towards rocksalt, according to the tendencies shown by both $\Delta_{so}$ and the conduction effective mass. This is in quite reasonable accordance to previous reports in the literature.

\begin{acknowledgments}
MEMR acnowledges support from Mexican CONACYT through Grant CB A1-S-8218. The authors thankfully acknowledge the computer resources, technical expertise and support provided by the Laboratorio Nacional de Superc\'omputo del Sureste de M\'exico, CONACYT, member of the network of national laboratories.
\end{acknowledgments}

\end{document}